\documentclass[12pt,preprint]{aastex}
\usepackage{epsf}
\usepackage{natbib}
\usepackage{float}
\usepackage{appendix}
\usepackage{amsmath}
\usepackage{pdflscape}

\tabletypesize{\scriptsize}

\newcommand{\rom}[1]{\uppercase\expandafter{\romannumeral #1\relax}}

\def\gappeq{\mathrel{ \rlap{\raise.5ex\hbox{$>$}}
                      {\lower.5ex\hbox{$\sim$}}  } }

\begin{document}
\shorttitle{Aperture Constraints for ExoEarth-Imaging Missions}
\shortauthors{}

\title{Lower Limits on Aperture Size for an ExoEarth-Detecting Coronagraphic Mission}

\author{Christopher C. Stark\altaffilmark{1}, Aki Roberge\altaffilmark{2}, Avi Mandell\altaffilmark{2}, Mark Clampin\altaffilmark{2}, Shawn D. Domagal-Goldman\altaffilmark{2}, Michael W. McElwain\altaffilmark{2}, Karl R. Stapelfeldt\altaffilmark{2}}

\altaffiltext{1}{Space Telescope Science Institute, 3700 San Martin Dr, Baltimore, MD 21218; cstark@stsci.edu}
\altaffiltext{2}{NASA Goddard Space Flight Center, Greenbelt, MD 20771}

\begin{abstract}

The yield of Earth-like planets will likely be a primary science metric for future space-based missions that will drive telescope aperture size.  Maximizing the exoEarth candidate yield is therefore critical to minimizing the required aperture.  Here we describe a method for exoEarth candidate yield maximization that simultaneously optimizes, for the first time, the targets chosen for observation, the number of visits to each target, the delay time between visits, and the exposure time of every observation.   This code calculates both the detection time and multi-wavelength spectral characterization time required for planets.  We also refine the astrophysical assumptions used as inputs to these calculations, relying on published estimates of planetary occurrence rates as well as theoretical and observational constraints on terrestrial planet sizes and classical habitable zones.  Given these astrophysical assumptions, optimistic telescope and instrument assumptions, and our new completeness code that produces the highest yields to date, we suggest lower limits on the aperture size required to detect and characterize a statistically-motivated sample of exoEarths.  

\end{abstract}


\section{Introduction}
\label{intro}

The direct detection of Earth-like planets will be a primary science goal for future space-based flagship missions and will likely drive telescope and instrument design, including aperture size.  Many studies have estimated the number of Earth-sized planets residing in the habitable zone (HZ) that could be detected with future missions \citep[e.g.,][]{brown2005,brownsoummer2010,savransky2010,turnbull2012,stark2014_2}.  However, these estimated yields vary by a factor of a few, even when adopting similar mission capabilities.  \citet{stark2014_2} showed this variation is due in part to differences in the exposure time distribution among candidate stars, and provided an Altruistic Yield Optimization (AYO) method to maximize exoEarth candidate yield by optimizing the exposure time distribution.  Remaining yield discrepancies are the result of differing astrophysical assumptions, observational requirements, and the ability of the yield calculation code to fully simulate the execution of a mission.

Here we build on the work of \citet{stark2014_2} to address these issues and update our calculations to include wavelength-dependent flux calculations, multiple visits per star, and spectral characterization time estimates.  We first address the astrophysical assumptions made when calculating exoEarth yield and provide suggested values derived from current published results.    In Section \ref{obs_assumptions}, we discuss our scientific motivation, as well as the observational, telescope, and instrument assumptions made.  In Section \ref{methods_section} we provide new yield optimization methods to incorporate multiple visits to each star, further maximizing exoEarth candidate yield.  We quantify the impact of these new methods and use them to estimate exoEarth candidate yield as a function of mission parameters in Section \ref{results_section}.  Finally, in Section \ref{discussion} we discuss our results, define a minimum scientifically compelling yield goal, and use this yield goal to provide rough constraints on the required aperture size.

\section{Assumptions \& Caveats}
\label{caveats}

\subsection{Astrophysical Assumptions}

The exoEarth candidate yield for a direct imaging mission is dependent on several assumed astrophysical quantities.  Most notably among these are the assumed size and optical properties for every Earth-sized planet residing in the habitable zone, the exozodiacal dust cloud surface brightness, and the number of habitable zone Earth-sized planets per star ($\eta_{\earth}$).  Here we briefly summarize our assumptions in each of these areas.

While certainly not all exoEarth candidates will be similar to Earth in size, composition, or albedo, we have no observational constraints on this variety.  Thus, most studies to date have simply adopted an Earth-twin as the definition of an exoEarth candidate, though the geometric albedo of this Earth-twin varies from study to study.  As in \citet{stark2014_2}, we continue to adopt an Earth-twin with a V band geometric albedo of $0.2$ as our definition of an exoEarth candidate.  Table \ref{exoearth_params_table} lists the parameters of our Earth-twin.

\begin{deluxetable}{ccl}
\tablewidth{0pt}
\footnotesize
\tablecaption{Baseline Astrophysical Parameters\label{exoearth_params_table}}
\tablehead{
\colhead{Parameter} & \colhead{Value} & \colhead{Description} \\
}
\startdata
$\eta_{\earth}$ & $0.1$ & Fraction of Sun-like stars with an exoEarth candidate \\
$R_{\rm p}$ & $1$ $R_{\earth}$ & Planet radius \\
$a$ & $[0.75,1.77]$ AU\tablenotemark{*} & Semi-major axis (uniform in $\log{a}$) \\
$e$ & $0$ & Eccentricity (circular orbits) \\
$\cos{i}$ & $[-1,1]$ & Cosine of inclination (uniform distribution) \\
$\omega$ & $[0,2\pi)$ & Argument of pericenter (uniform distribution) \\
$M$ & $[0,2\pi)$ & Mean anomaly (uniform distribution) \\
$\Phi$ & Lambertian & Phase function \\ 
$A_G$ & $0.2$ & Geometric albedo of planet at $0.55$ and $1$ $\mu$m \\
$z$ & 23 mag arcsec$^{-2}$\tablenotemark{\dag}  & V band surface brightness of zodiacal light \\
$x$ & 22 mag arcsec$^{-2}$\tablenotemark{\ddag}  & V band surface brightness of 1 zodi of exozodiacal dust \\
$n$ & $3$ & Number of zodis for all stars \\
\enddata
\vspace{-0.1in}
\tablenotetext{*}{$a$ given for a solar twin.  The habitable zone is scaled by $\sqrt{L_{\star}/L_{\sun}}$ after calculating projected separation $s_{\rm p}$.}
\tablenotetext{\dag}{Varies with ecliptic latitude---see Appendix B in \citet{stark2014_2}.}
\tablenotetext{\ddag}{For Solar twin. Varies with spectral type---see Appendix C in \citet{stark2014_2}.}
\end{deluxetable}

We assume the same definition of 1 ``zodi" of exozodiacal dust as \citet{stark2014_2}: a fixed optical depth that produces a V band surface brightness of 22 mag arcsec$^{-2}$ around a Solar twin (see Appendix C.2 in \citet{stark2014_2}), but also calculate the wavelength-dependent flux.  We also adopt the same local zodi flux calculations (see Appendix B in \citet{stark2014_2}), but include the wavelength dependence of the zodiacal cloud.  In Section \ref{spec_char_time_section}, we discuss our calculation of the wavelength dependence of both of these sources.

The most significant change to our astrophysical assumptions relates to our definition of the classical habitable zone (HZ).  \citet{stark2014_2} adopted the \citet{brown2005} HZ definition (BHZ).  \citet{brown2005} distributed orbital semi-major axes from 0.7 -- 1.5 AU for a Sun-like star, scaling each semi-major axis by $\sqrt{L_{\star}/L_{\sun}}$.  Detailed estimates of the classical HZ are slightly more distant.  \citet{kopparapu2013} presents several limits to the inner and outer edges of the classical HZ, with an ``optimistic HZ" (OKHZ) ranging from 0.75 -- 1.77 AU and a ``pessimistic HZ" (PKHZ) ranging from 0.99 -- 1.67 AU for a Sun-like star.  \citet{kopparapu2013} shows that the classical HZ deviates by a few percent from the $\sqrt{L_{\star}/L_{\sun}}$ scaling relation, though we continue use of this simple scaling relation out of expediency.

Additionally, \citet{brown2005} distributed orbits uniformly in semi-major axis and eccentricity from 0 -- 0.35.  As a result, the exoEarth candidates in \citet{brown2005} can range in circumstellar distance from roughly 0.5 -- 2.0 AU for a Sun-like star.  We adopt semi-major axes distributed uniformly in logarithmic space, as supported by radial velocity surveys and \emph{Kepler} observations \citep[e.g.,][]{cumming2008,howard2012,petigura2013}.  Though we have little direct evidence of the eccentricities of HZ Earth-sized planets, \citet{kane2012} showed a significant trend in the average eccentricity derived from Kepler data.  By comparing deviations in the transit durations as a function of planet radius to model orbits, \citet{kane2012} showed that planets smaller than Neptune have increasingly circular orbits, likely due to the fact that smaller planets tend to be in multi-planet systems.  We therefore adopt circular orbits.

\begin{figure}[H]
\begin{center}
\includegraphics[width=4in]{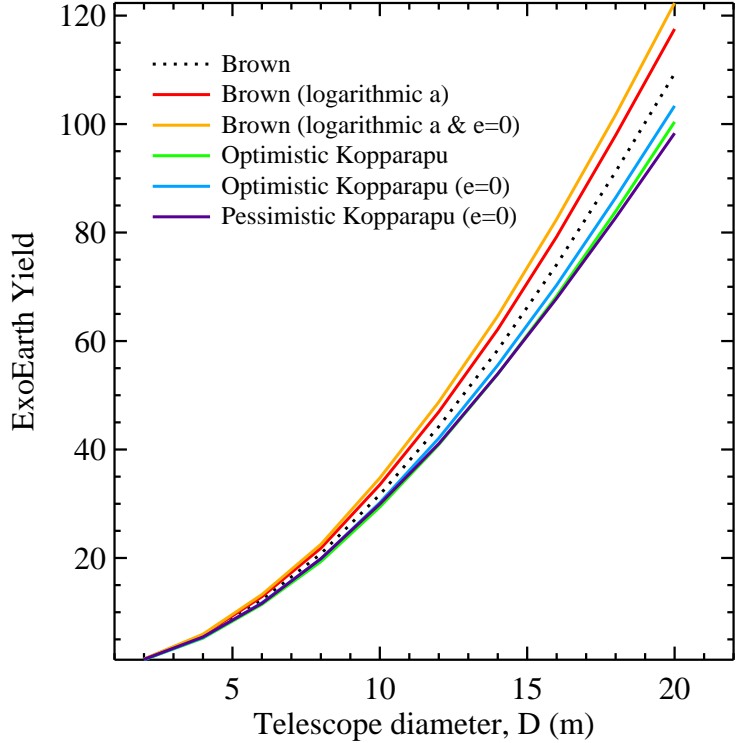}
\caption{Sample yields as functions of telescope diameter for six different distributions of HZ orbits, assuming a fixed valued of $\eta_{\earth}$. The assumed orbital distribution alone can produce a $\sim20\%$ spread in yield. \label{HZyield_fig}}
\end{center}
\end{figure}

The assumed orbital distribution of planets can lead to differences in expected yield.  For illustrative purposes, Figure \ref{HZyield_fig} shows example yields as functions of aperture size for six different assumptions about the HZ orbits, assuming a fixed value of $\eta_{\earth}$.  These different definitions result in a $\sim20\%$ spread in yield.

However, each HZ definition should have its own corresponding value for $\eta_{\earth}$.  Recent estimates of $\eta_{\earth}$ for Solar type stars suggest $\eta_{\earth} \approx 0.2$ for planets with radii 1--2 $R_{\earth}$ receiving 0.25 -- 4 times the Earth's incident Solar flux \citep{petigura2013_2,silburt2014}.  \citet{petigura2013_2} also found that $\eta_{\earth}$ scales with the area of the box with dimensions $\log{(R_{\rm planet, max} / R_{\rm planet, min})}$ $\times$ $\log{(a_{\rm outer\, edge} / a_{\rm inner\, edge})}$.  Thus, to calculate a value for $\eta_{\earth}$ for Sun-like stars using the results of \citet{petigura2013_2} and \citet{silburt2014}, we must also define the range of planet radii that we consider ``Earth-like."

Estimates of the bulk densities of planets as a function of radius show a transition at $\sim1.5$ $R_{\Earth}$ \citep[e.g.,][]{weiss2014,wolfgang2014}.  Planets smaller than $\sim1.5$ $R_{\Earth}$ are likely rocky in composition while larger planets are likely enveloped by a H$/$He atmosphere.  We therefore adopt $1.5$ $R_{\Earth}$ as the maximum exoEarth candidate radius.  

Although there is no firm lower-limit on the size of a potentially habitable world, loose guidelines can be derived from the work of \citet{zahnle2013}. This work proposes a lower limit on the mass of habitable planets given the input of high energy radiation to the planet and theoretical atmospheric loss processes.  For a planet that receives the same flux from its star that Earth receives from the Sun, the lower limit is $\sim0.6$ $R_{\Earth}$ assuming an Earth-like bulk density.

However, we also require that our Earth-twin assumption adopted for yield calculations be representative of the typical exoEarth candidate radius.  Thus, we demand that the median planet radius be equal to 1 $R_{\Earth}$.  This results in our adopted minimum radius of $0.66$ $R_{\Earth}$, roughly consistent with the minimum radius suggested by \citet{zahnle2013}.  Given this range of planet radii, we scaled $\eta_{\Earth}$ from the results of \citet{petigura2013_2} for each of the three classical HZ definitions discussed, maintaining a constant fractional uncertainty.  The results are presented in Table \ref{eta_Earth_table}.

\begin{deluxetable}{llccc}
\tablewidth{0pt}
\footnotesize
\tablecaption{$\eta_{\Earth}$ for Planet Radii 0.66 -- 1.5 $R_{\earth}$ \label{eta_Earth_table}}
\tablehead{
\colhead{HZ Definition} & \colhead{Acronym} & \colhead{$a_{\rm inner}$\tablenotemark{1}} & \colhead{$a_{\rm outer}$\tablenotemark{1}} & \colhead{$\eta_{\earth}$\tablenotemark{2}}\\
 & & \colhead{(AU)} & \colhead{(AU)} \\
}
\startdata
\citet{brown2005} & BHZ & $0.7$ & $1.5$ & $0.14\pm0.05$ \\
Optimistic \citet{kopparapu2013} & OKHZ & $0.75$ & $1.77$ & $0.16\pm0.06$ \\
Pessimistic \citet{kopparapu2013} & PKHZ & $0.99$ & $1.67$ & $0.10\pm0.04$ \\
\enddata
\tablenotetext{1}{For a Sun-like star.}
\tablenotetext{2}{Based on results and scaling relations from \citet{petigura2013_2}.}
\end{deluxetable}

\begin{figure}[H]
\begin{center}
\includegraphics[width=4in]{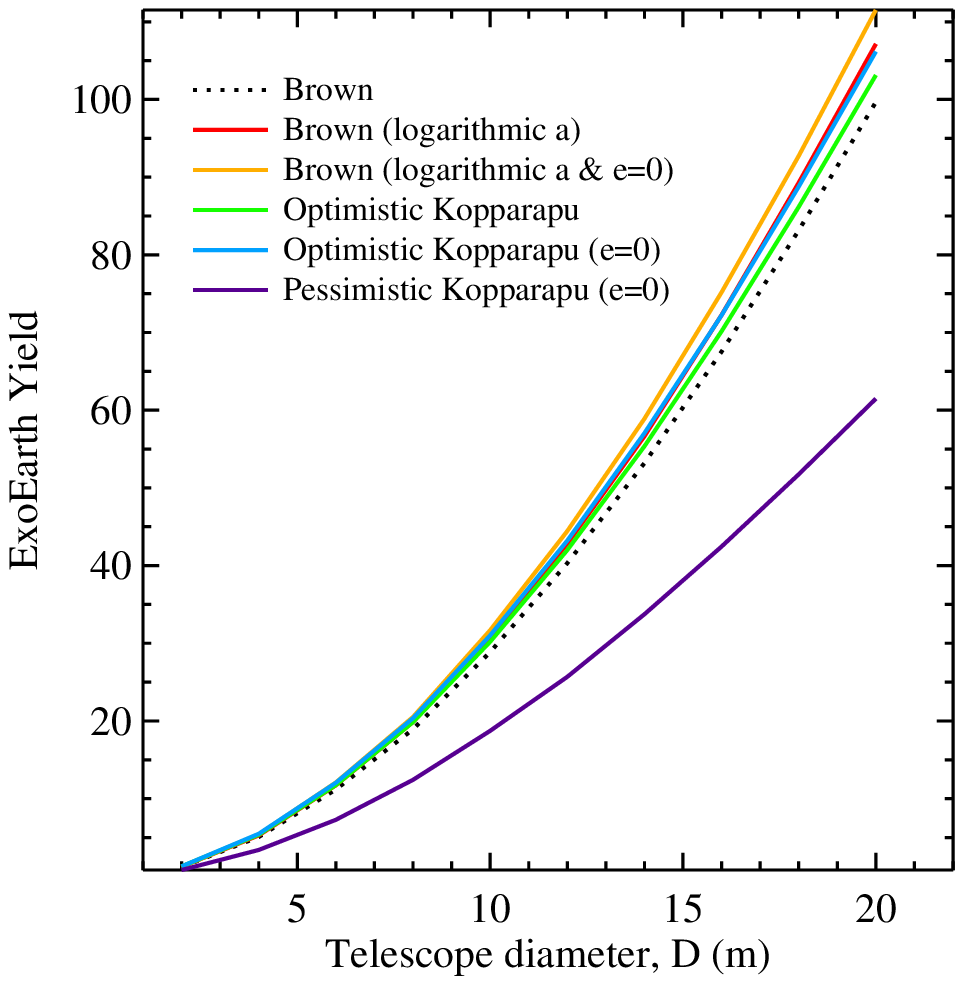}
\caption{Sample yields as functions of telescope diameter for six different distributions of HZ orbits, assuming the self-consistent values of $\eta_{\earth}$ listed in Table \ref{eta_Earth_table}. \label{HZyield2_fig}}
\end{center}
\end{figure}

Figure \ref{HZyield2_fig} shows the same sample yields shown in Figure \ref{HZyield_fig}, but scaled by the $\eta_{\Earth}$ values listed in Table \ref{eta_Earth_table}.  The spread in the yield between the BHZ and the OKHZ definitions decreases to $\sim10\%$.  The yield associated with the PKHZ is $\sim40\%$ smaller.

The choice of which classical HZ to adopt is a judgement call.  Designing a mission around a pessimistic HZ will lead to a larger required aperture and will more finely tune the mission to a smaller region of orbital space.  Choosing an optimistic HZ may result in more Earth-sized planets with a smaller aperture, but fewer truly Earth-like planets.  We note that the 1$\sigma$ lower limit on $\eta_{\Earth}$ for the OKHZ is equal to the expected value of $\eta_{\earth}$ for the PKHZ.  \emph{For the purposes of this paper, we choose the OKHZ definition with planets on circular orbits distributed uniformly in the logarithm of semi-major axis, and use the conservative 1$\sigma$ lower limit on $\eta_{\Earth}$ of 0.1.}

We based our $\eta_{\earth}$ assumption on the work of \citet{petigura2013_2}.  However, other recent planet occurrence rate estimates suggest somewhat smaller values of $\eta_{\earth}$ \citep[e.g.,][]{dong2013,foremanmackey2014}.  Even when adopting the same catalog as \citet{petigura2013_2}, \citet{foremanmackey2014} derives an $\eta_{\earth}$ value three times smaller than that of \citet{petigura2013_2}.  We therefore do not regard our adopted value of $\eta_{\earth} = 0.1$ as pessimistic.

\subsection{Telescope, Instrument, \& Observational Assumptions}
\label{obs_assumptions}

We calculate exoEarth candidate yield for a coronagraphic mission only, ignoring alternative high-contrast imaging technologies like an external starshade.  We calculate yield via Monte Carlo simulations of completeness.  We ignore scheduling constraints and the effects of any real-time decisions that may be made during the course of the mission.

We also ignore the impacts of overheads associated with telescope slews and wavefront control, which are architecture-dependent.  The overheads for a coronagraph may be as long as the observations themselves.  Thus, in reality the required mission lifetime may be twice as long as the assumed total exposure time of the simulations in this work.  Additionally, pointing overheads for the telescope may significantly reduce the yields of the largest apertures we simulate, for which the slew time may be comparable to some of the exposure times.  We leave the addition of overheads to future work.

We assume a mission that has the primary goal of detecting an Earth-\emph{like} planet.  The mission must therefore be capable of remotely characterizing any exoEarth candidates to determine whether or not they are Earth-like.  There is no agreed-upon definition of ``Earth-like," but one commonly accepted minimum requisite is liquid surface water, potentially resulting in detectable amounts of atmospheric water vapor.  The water vapor feature that is commonly viewed as the most readily detectable occurs near $0.95$ $\mu$m \citep[e.g.,][]{brandt2014}.  We therefore adopt the search for atmospheric water vapor at 1 $\mu$m as an observational requirement.  While we continue to assume V band detections of exoEarth candidates, \emph{only the exoEarth candidates directly observable at a wavelength of $1$ $\mu$m count toward the total yield of the mission.}

We have updated the code of \citet{stark2014_2} to include the time required for spectral characterization of all detected exoEarth candidates at 1 $\mu$m.  The spectral characterization time counts against the total exposure time budget of the mission, i.e. a higher detection rate results in more time spent characterizing.  For the baseline mission, we require an $R=50$ spectrum with S/N $=5$ per spectral resolution element at a wavelength of 1 $\mu$m for every detectable exoEarth candidate, which provides a 50\% probability of a 3$\sigma$ detection of water vapor \citep{brandt2014}.  We implicitly assume that all non-exoEarth candidates (all planets outside of the HZ and all planets within the HZ that are not Earth-sized) can be distinguished from exoEarth candidates without taking $R=50$ spectra, either by brightness, color information, or lower resolution spectra that require significantly less exposure time.  

The search for water defines our minimum characterization requirement.  Searching for additional biosignatures may require additional exposure time if multi-band observations cannot be made in parallel using multiple coronagraphs.  However, this additional time could be reduced by applying a ``ladder of biosignatures," in which only the exoEarth candidates that exhibit detectable traces of water are followed up for additional characterization.

Unlike \citet{stark2014_2}, we include the effects of detector noise.  We assume noiseless detectors for the baseline mission and address the impact of detector noise in Section \ref{sensitivity_section}.  We also approximately include coronagraphic broadening of the planet's PSF via the broadening factor $b$, which linearly scales the area of the planet's Airy pattern PSF.  We assume a diffraction-limited planet PSF ($b=1$) for the baseline mission.

Finally, we allow stars to be observed multiple times, i.e. revisited, to increase the chances of detecting exoEarth candidates.  We place no restrictions on the exposure time of each observation.  We do not include any revisits that may be required to establish the orbit of a potentially habitable planet.  Depending on the observation plan implemented, orbit-constraining revisits may be rarely required, and we expect their impact on the overall mission yield to be small given that most stars are already assumed to be observed multiple times for the purposes of increasing the yield.

\begin{deluxetable}{ccl}
\tablewidth{0pt}
\footnotesize
\tablecaption{Baseline Mission Parameters\label{baseline_params_table}}
\tablehead{
\colhead{Parameter} & \colhead{Value} & \colhead{Description} \\
}
\startdata
$D$ & $10$ m & Telescope diameter \\
$\Sigma \tau$ & $1$ yr & Total exposure time of the mission \\
$T$ & $0.2$ & End-to-end facility throughput \\
$X$ & $0.7$ & Photometry aperture radius in $\lambda/D$ \\
$\Upsilon$ & $0.69$ & Fraction of Airy pattern contained within photometry aperture \\
$\Omega$ & $b\pi(X\lambda/D)^2$ radians & Solid angle subtended by photometry aperture \\
$b$ & $1.0$ & Areal broadening of the planet's PSF \\
IWA & $3.6\lambda_{\rm d}/D$ (52 mas) & Inner working angle for both coronagraphs \\
OWA & $15\lambda_{\rm d}/D$ (213 mas) & Outer working angle for both coronagraphs \\
$\Delta$mag$_{\rm floor}$ & $2.5 - 2.5\log{\zeta_{\rm d}}$ & Systematic noise floor for both coronagraphs (i.e., dimmest point source detectable at S/N)\\

\hline
$\lambda_{\rm d}$ & $0.55$ $\mu$m & Central wavelength for detection (V band) \\
$\Delta \lambda_{\rm d}$ & $0.11$ $\mu$m & Bandwidth for detection \\
S/N$_{\rm d}$ & $7$ & Signal to noise ratio required for broadband detection of a planet \\
$\zeta_{\rm d}$ & $10^{-10}$ & Raw contrast level in detection region, relative to theoretical Airy pattern peak \\
${\rm CR}_{\rm b,detector,d}$ & $0$ & Detector noise count rate for detection \\
\hline

$\lambda_{\rm c}$ & $1.0$ $\mu$m & Central wavelength for spectral characterization \\
$R_{\rm c}$ & $50$ & Spectral resolving power required for spectral characterization \\
S/N$_{\rm c}$ & $5$ & Signal to noise ratio per spectral resolution element required for spectral characterization \\
$\zeta_{\rm c}$ & $5\times10^{-10}$ & Raw contrast level for spectral characterization, relative to theoretical Airy pattern peak \\
${\rm CR}_{\rm b,detector,c}$ & $0$ & Detector noise count rate for spectral characterization\\

\enddata
\end{deluxetable}

Table \ref{baseline_params_table} lists the baseline telescope and instrument parameters.  We assume a telescope of diameter $D$ paired with two coronagraphs, each with its own detector, whose parameters are chosen optimistically to represent future performance.  The first coronagraph is designed for broadband detection at $\lambda_{\rm d}=0.55$ $\mu$m.  We will refer to this hereafter as the ``detection coronagraph" and subscript its defining parameters with a `d.' Because $\eta_{\earth}=0.1$ implies a $<10\%$ chance of detecting an exoEarth candidate during any observation, many observations must be made, motivating the need for an optimistic raw contrast of $10^{-10}$ to reduce detection exposure times.  

The second coronagraph is designed for spectral characterization at $\lambda_{\rm c}=1$ $\mu$m.  We will refer to this hereafter as the``characterization coronagraph" and subscript its defining parameters with a `c.' Because the inner working angle of a coronagraph roughly scales with $\lambda/D$ and we require the detection of water for every exoEarth candidate, the inner working angle of our 1 $\mu$m characterization coronagraph defines which planets are observable.  We thus choose the same physical IWA for both coronagraphs, with an optimistic IWA$_{\rm c} = 2\lambda_{\rm c}/D$ and a relaxed IWA$_{\rm d}=3.6\lambda_{\rm d}/D$.  Because the two inner working angles are the same physical size, we will refer to them simply with the singular parameter IWA.  

Realistically, the optimistic IWA$_{\rm c} = 2\lambda_{\rm c}/D$ may have to come at the expense of contrast.  Because we assume spectral characterization only occurs for true exoEarth candidates, spectral characterization will occur on relatively few targets.  Thus, the characterization time may not greatly impact the overall yield.  In light of this we relax the baseline raw contrast requirement at 1 $\mu$m to $\zeta_{\rm c}=5\times10^{-10}$.  Both coronagraphs are also assumed to have the same physical OWA as well as the same absolute noise floor.  \emph{In summary, we adopt a coronagraph for 1 $\mu$m spectral characterization that is optimized for small IWA, and a coronagraph for V band detection that is optimized for high contrast/short exposure times.}

We assume a total exposure time of 1 year out of a 5 year mission (though we do not explicitly budget for overheads, 1 additional year may be required for telescope and coronagraph overheads).  We assume the faintest point source that can be detected at the threshold S/N is $2.5$ magnitudes fainter than the raw detection contrast $\zeta_{\rm d}$, i.e. post-processing can reveal point sources with relative flux equal to one tenth that of the raw detection contrast, and one fiftieth that of the raw spectral characterization contrast for the baseline values of $\zeta_{\rm d}$ and $\zeta_{\rm c}$.  Second, we relax the threshold S/N for detection from a value of 10 to 7, a choice we explain in Section \ref{sensitivity_section}.

We also make several minor changes to the root target list.  In \citet{stark2014_2}, we selected targets within 50 pc from the \emph{Hipparcos} catalog, removing only those targets with known companions within $10\arcsec$.  Here, we also remove stars categorized as luminosity classes \rom{1}, \rom{2}, and \rom{3} as well as stars that have no classification, to select only main sequence and sub-giant stars.  These changes are negligible, impacting the calculated yield at the $1\%$ level.

We note that we did not vet the input target list in great detail.  We vetted binaries by simply removing those stars with companion separations $<10\arcsec$, where companion separations were taken as the minimum of the separations reported in the \emph{Hipparcos} Double \& Multiples Catalog and the Washington Double Star Catalog.  Additionally, the \emph{Hipparcos} catalog is missing several nearby, bright stars that would make excellent targets for exoEarth searches.  A more thorough vetting of candidates would reference additional catalogs for missing stars, remove any additional bad targets from the \emph{Hipparcos} input list, and estimate future binary configurations.  However, under the assumption of $\eta_{\earth} = 0.1$, 10 additional/fewer high priority targets with fully observable habitable zones would change the yield by less than 1 exoEarth candidate, so we expect the impact of these additional screenings on the mission yields reported in this paper to be minimal.

\section{Methods}
\label{methods_section}

The fundamentals of our exoEarth candidate yield calculations are based on \citet{brown2005} and described in \citet{stark2014_2}.  Here we discuss updates to the code.

\subsection{Detector Noise}

\citet{stark2014_2} calculated the required exposure time assuming negligible detector noise.  Here we explicitly include the detector noise count rate as an additional parameter to constrain acceptable levels.  We calculate the background count rate as
\begin{equation}
	{\rm CR}_{\rm b} = {\rm CR}_{{\rm b,}\star} + {\rm CR}_{\rm b,zodi} + {\rm CR}_{\rm b,exozodi} + {\rm CR}_{\rm b,detector},
\end{equation}
where ${\rm CR}_{{\rm b,}\star}$, ${\rm CR}_{\rm b,zodi}$, and ${\rm CR}_{\rm b,exozodi}$ are the count rates of the unsuppressed stellar light, local zodi, and exozodi, given by Eq. 11, 13, and 15 in \citet{stark2014_2}, respectively.  The additional term, ${\rm CR}_{\rm b,detector}$, represents the noise count rate of the detector, a parameter we control.  We can relate read noise ${\rm RN}$ and dark count rate $\xi$ to ${\rm CR}_{\rm b,detector}$ via
\begin{equation}
 \label{detector_noise_eq}
 	{\rm CR}_{\rm b,detector} \approx n_{\rm pix} \left(\xi + {\rm RN}^2 / \tau_{\rm expose} \right),
\end{equation}
where $n_{\rm pix}$ is the number of pixels required for imaging or spectral characterization and $\tau_{\rm expose}$ is the length of an individual exposure.  In practice, we assume the detection and characterization coronagraphs have independent detectors and allow for corresponding values of ${\rm CR}_{\rm b,detector,d}$ and ${\rm CR}_{\rm b,detector,c}$, respectively.  The above expressions underestimate the read noise contribution for observation times $\tau \lesssim \tau_{\rm expose}$, but this only occurs for observations of bright targets, for which the unsuppressed starlight will dominate the noise sources.

\subsection{Spectral Characterization Time}
\label{spec_char_time_section}

We calculate the required spectral characterization time, $t_{\rm c}$, for every synthetic exoEarth around every star using the same basic exposure time expressions given in Section 3 of \citet{stark2014_2}.  However, we use the updated expression for ${\rm CR}_{\rm b}$ given above, substitute S/N$_{\rm c}$ for S/N, and substitute $\Delta\lambda_{\rm c}=\lambda_{\rm c}/R_{\rm c}$ for $\Delta\lambda$.

We also adjust the flux of every source to the desired spectral characterization wavelength of 1 $\mu$m.  We assume a wavelength-independent geometric albedo for Earth, such that the planet's flux is linearly proportional to the stellar spectrum.  This assumption is valid for this work, as the Earth's continuum geometric albedo is approximately $0.2$ between $0.5$ and $1.2$ $\mu$m \citep{robinson2011}.  We calculate the flux of each star at 1 $\mu$m by linearly extrapolating the stellar fluxes from the $V$ and $I$ magnitudes given in the \emph{Hipparcos} catalog.  Although this extrapolation to 1 $\mu$m is a rough estimate, it suffices for our calculations, as the yield is a weak function of total exposure time and thus negligibly impacted by modest changes to the spectral characterization time \citep{stark2014_2}.

For the local zodi, we calculate the wavelength dependence by interpolating Table 19 of \citet{leinert1998}.  We calculate the 1 $\mu$m brightness of the exozodi and flux of the planet around each star by multiplying the V band flux of each source by $F_{\star}(1\, \mu\rm{m})/F_{\star}(0.55\, \mu\rm{m})$, where $F_{\star}(1\, \mu\rm{m})$ and $F_{\star}(0.55\, \mu\rm{m})$ are the stellar fluxes at 1 $\mu$m and $0.55$ $\mu$m, respectively.  

Given that the spectral characterization time will count against the total exposure time budget, such a mission cannot afford to take $R_{\rm c}=50$, S/N$_{\rm c}=5$ spectra of every system.  However, it would be risky to wait to take spectra, as the planet could disappear interior to the IWA, move into crescent phase, or become confused with another object.  Thus, the choice to spectrally characterize a planet must be made in near-real time.

The most robust treatment of spectral characterization therefore requires some sort of Monte Carlo mission execution simulator, run many times over to statistically bound the impact of spectral characterization time.  Such a simulation would be computationally expensive compared to our yield estimation code and is more appropriate for modeling a small number of mission designs.  In lieu of robust mission execution simulations, we include the impact of spectral characterization time in a probabilistic sense that approximates a large suite of Monte Carlo simulations.

To approximate the required spectral characterization time of the mission, we require a spectrum for every \emph{detected} exoEarth candidate.  We start by calculating $t_{\rm c}$ for every synthetic exoEarth simulated around each star.  We calculate the probabilistic spectral characterization time for the $i^{\rm th}$ star as
\begin{equation}
	\label{tchar_eq}
	\tau_{i,{\rm c}} = \eta_{\earth}\; C_i\, \langle t_{\rm c} \rangle,
\end{equation}
where $C_i$ is the completeness of the $i^{\rm th}$ star, the product $\eta_{\earth} C_i$ is the probability of detecting an exoEarth candidate around the $i^{\rm th}$ star, and $\langle t_{\rm c} \rangle$ is the average spectral characterization time of all detectable synthetic exoEarths around the $i^{\rm th}$ star.  Because 
\begin{equation}
	C_i = \frac{n_{i,{\rm detectable}}}{n_{\rm simulated}},
\end{equation}
where $n_{i,{\rm detectable}}$ is the number of detectable synthetic exoEarths around the $i^{\rm th}$ star and $n_{\rm simulated}$ is the total number of simulated synthetic exoEarths around each star, and 
\begin{equation}
	\langle t_{\rm c} \rangle = \sum\limits_{j=1}^{n_{i,{\rm detectable}}} \frac{t_{\rm c}}{n_{i,{\rm detectable}}},
\end{equation}
we can rewrite Equation \ref{tchar_eq} as
\begin{equation}
	\tau_{i,{\rm c}} = \frac{\eta_{\earth}}{n_{\rm simulated}} \sum\limits_{j=1}^{n_{i,{\rm detectable}}} t_{j,{\rm c}},
\end{equation}
where the quantity $\eta_{\earth}/n_{\rm simulated}$ is the probability of observing an individual simulated synthetic exoEarth and the summation is simply the total of the spectral characterization times for all detectable synthetic exoEarths.  We note that the spectral characterization time counts against the mission's total exposure time budget; missions that detect more planets must devote the required spectral characterization time to them.

\subsection{Optimized Revisits}

The majority of completeness calculations thus far have assumed a single visit to each star, the most notable exception being \citet{brownsoummer2010}, who include revisits via an analytic approximation.  The single-visit assumption is commonly made for two reasons.  First, the initial visit to a target star is the most productive visit, since no portion of orbital space has been ruled out.  Strictly speaking, any revisit can only achieve a maximum completeness of one minus the previous visits' total completeness \citep{brownsoummer2010}.  However, an additional visit to a high priority star may prove more fruitful than the first visit to a lower priority star.

Second, revisits require computationally intensive calculations.  The orbit of every unobserved synthetic planet must be advanced by the delay time between revisits, $\Delta t$, prior to calculating the revisit completeness.  Further, one would ideally optimize the delay time $\Delta t$ to make the revisit as efficient as possible.  \citet{brownsoummer2010} estimated that the revisit completeness of a single star calculated for a single delay time requires on the order of a few seconds on a single processor; optimizing the delay time for hundreds of stars, each with multiple visits, could take hundreds of hours.  In light of this, \citet{brownsoummer2010} opted for a non-optimized approximation to the revisit completeness.

We have overcome the numerical run time limitations of revisits, enabling optimization.  We accomplished this by indexing the orbits of every simulated planet in mean anomaly, avoiding on-the-fly orbital calculations, and prioritizing revisit calculations to avoid unnecessary computations.  We can simultaneously optimize the number of revisits to each star, the delay time between revisits, and the exposure time of every visit to maximize mission yield in a reasonable run time.  For example, simulating $10^5$ synthetic exoEarths around every star, our code can optimize up to 10 visits to thousands of potential target stars in $\sim5$ minutes on a single processor.

To calculate revisits, we divide each synthetic exoEarth orbit into $n_{M}$ steps in mean anomaly.  Each of these steps in mean anomaly counts as a potentially observable planet.  For example, to simulate $10^5$ synthetic exoEarths, we sample $10^3$ orbits 100 times each, i.e. $n_{\rm planets} = n_{\rm orbits} \times n_{M}$.

For each star, we calculate the projected separations and required exposure time of all $n_{\rm planets}$ planets in advance.  At any future time, we can then calculate a given planet's evolved projected separation and exposure time
by referencing the mean anomaly index that most closely corresponds to a delay time of $\Delta t$ via
\begin{equation}
	M\!\left(\Delta t\right) = n \; \Delta t - M\!\left(t=0\right),
\end{equation}
where $n$ is the mean motion.  To calculate the mean motion of every orbit, we make the approximation $(M_{\star} / M_{\sun}) \approx (L_{\star} / L_{\sun})^{0.28}$, where $M_{\star}$ and $L_{\star}$ are the stellar mass and luminosity, respectively.  We note that the validity of this scaling does not impact any of the results presented here, and will only become important when determining a real world observing schedule.

We then calculate the first visit completeness as a function of exposure time as described in \citet{stark2014_2}.  Once the completeness, $C_1$, and exposure time, $\tau_1$, for the first visit is chosen by optimizing the first visit exposure time, we follow these steps to optimize the $i^{\rm th}$ visit to a given star:
\begin{enumerate}
	\item Mark all planets with IWA $< s_p(\Delta t_{i-1}) <$ OWA and $\tau_p(\Delta t_{i-1}) < \tau_{i-1}$ as having been observed, where $s_p$ and $\tau_p$ are the projected separation and exposure time for planet $p$.
	\item Advance all unobserved planets by $\Delta t_i$.
	\item Select all advanced, unobserved planets with IWA $< s_p(\Delta t_{i}) <$ OWA and sort them by $\tau_p(\Delta t_{i})$
	\item Calculate $C_i(\tau_i)$ and $dC_i/d\tau_i$.
	\item Optimize $\tau_i$ (see below).
	\item Repeat steps 2--5 for multiple values of $\Delta t_i$.  Choose $\Delta t_i$ such that $C_i / \tau_i$ is maximized, thereby optimizing both $\Delta t_i$ and $\tau_i$.
\end{enumerate}

A range of visit delay times $\Delta t_i$ is chosen that allows planets to evolve in their orbits without losing phase coherence.  Ideal values of $\Delta t_i$ should correspond to the time for unobserved crescent phase planets to evolve to gibbous phase or planets behind the IWA to become visible.  Thus, the optimal value of $\Delta t_i$ is a fraction of the median orbital period.  In practice, we find 20 uniformly spaced values of $\Delta t_i$ from 0 to the median orbital period well resolves the optimum delay time.

The procedure above requires optimizing the exposure time of every visit, including the first.  The AYO algorithm introduced in \citet{stark2014_2} provides a numerically efficient method for optimizing the exposure time of every star in the single-visit case by redistributing ``packets" of exposure time among stars.  However, because the completeness of each revisit is a function of the planets left unobserved from previous visits, redistributing a single ``packet" of exposure time from one star to another requires recalculating $C_i(\tau_i)$ for all subsequent visits.  This in principle could be done, but would likely be computationally inefficient, since calculating $C_i(\tau_i)$ dominates the run time of the code. 

Instead we opt for the alternative ``equal-slope" method introduced by \citet{hunyadi2007}.  Briefly, the equal-slope method works under the assumption that redistributing ``packets" of exposure time to stars with the largest $dC/d\tau$ will result in all selected stars having equal values of $dC/d\tau$.  One must therefore guess at the optimal value of $dC/d\tau$, solve for the $\tau_i$ at which $dC_i/d\tau_i = dC/d\tau_{\rm guess}$ for each star, then down-select to the set of observations that maximize the total completeness and fit within the total exposure time of the mission.  This is repeated until the optimized value of $dC/d\tau$ is found, a process that typically converges in $\sim10$ iterations.

\begin{figure}[H]
\begin{center}
\includegraphics[width=6in]{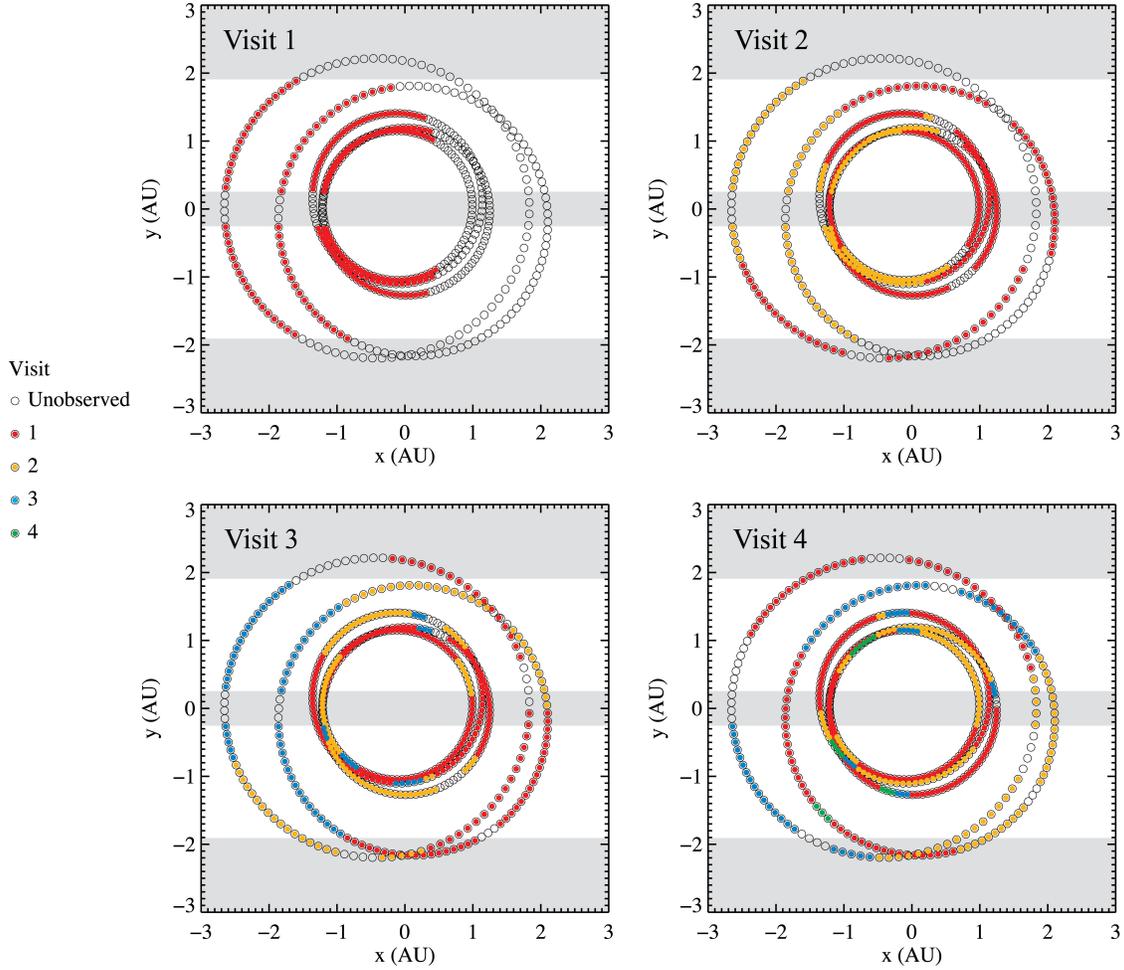}
\caption{\emph{Top-left}: Five sample orbits around HIP 27072 seen edge-on by an 8 m telescope at $x=+\infty$, resolved into 100 mean anomaly steps. IWA and OWA shown in gray.  Planets detectable during the first visit, primarily in gibbous phase, are shown in red.  \emph{Top-right}: First visit planets evolved, detectable second visit planets shown in orange.    \emph{Bottom-left:}  Planets continue to evolve as third visit is made (blue).  \emph{Bottom-right:} Detectable planets after four visits. \label{revisit_illustration_fig}}
\end{center}
\end{figure}

Figure \ref{revisit_illustration_fig} illustrates the revisit method at work for our baseline mission observing HIP 27072, an F6V star at 9 pc with a $\sim$1000 AU K2V companion.  For this illustration, we have used the BHZ with eccentric orbits allowed.  The telescope is assumed to be at $x=+\infty$, pointed to the left.  The gray shaded regions mark the IWA and OWA, interior and exterior to which planets cannot be detected, respectively.  We select five sample orbits, all seen edge-on by the telescope, that are resolved to 100 steps in mean anomaly.  Planets with positive values of $x$ appear to be crescent phase, while those with negative values appear to be in gibbous phase.

In the top-left panel, planets that could be observed during the first visit, mostly in gibbous phase, are marked with red dots.  Between the IWA and OWA, the transition between red and open circles traces a contour of constant planet brightness.  In the top-right panel, the planets have evolved in their orbits for a time equal to $\Delta t_2$, at which time the most unobserved planets appear in gibbous phase, triggering the second visit (orange).  The full simulation\footnote{Available at www.starkspace.com/revisits.mp4} ends in the bottom right panel, which shows a total of four visits calculated for this star.

\begin{figure}[H]
\begin{center}
\includegraphics[width=4in]{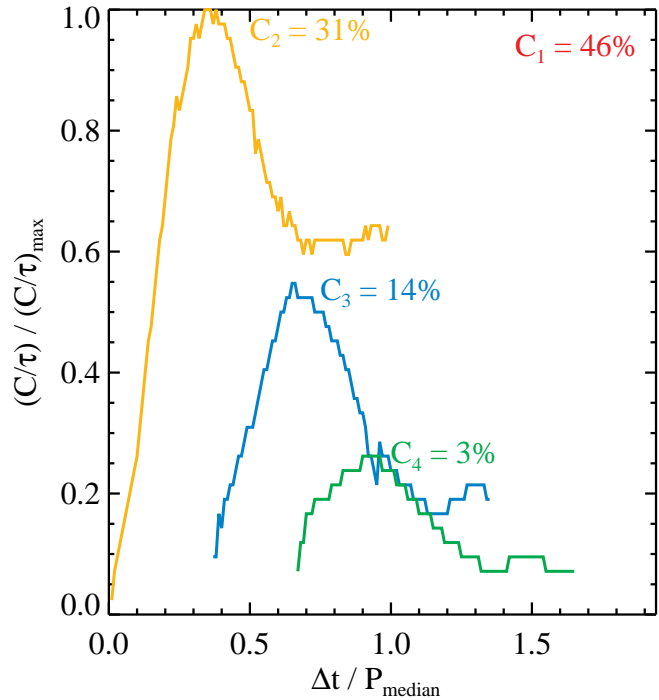}
\caption{Benefit-to-cost ratio as a function of delay time for the four visits selected for HIP 27072.  Completeness attained for each visit is indicated near the peak cost-to-benefit ratio.  Optimizing the delay and exposure time of revisits can double the benefit-to-cost ratio.  \label{revisit_plot_fig}}
\end{center}
\end{figure}

Figure \ref{revisit_plot_fig} shows the calculated benefit-to-cost ratio for this set of observations.  The first visit completeness $C_1 = 0.46$.  The second through fourth visit benefit-to cost ratio are plotted as functions of delay time.  The code selects the delay time at which the benefit-to-cost ratio is maximized, to achieve the most completeness for as little exposure time as possible.  Each peak is labeled with the value of completeness obtained at that time.  We obtain 94\% completeness in four visits to this star.

\section{Results}
\label{results_section}

\subsection{The impact of optimized revisits}

\begin{figure}[H]
\begin{center}
\includegraphics[width=6in]{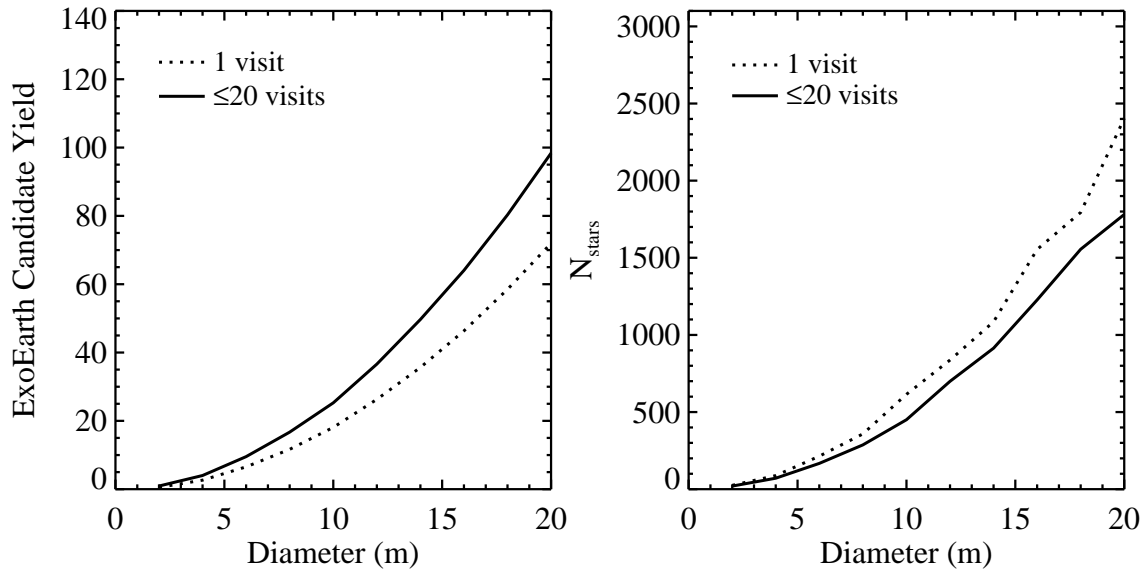}
\caption{Comparison of ExoEarth candidate yield (left) and number of unique stars observed (right) as functions of aperture size for the single visit and multi-visit cases.  No spectral characterization time is included in these calculations.  \label{revisits_impact}}
\end{center}
\end{figure}

\citet{stark2014_2} did not include revisits.  Here we show the change in yield when we ``turn on" revisits, but do not include spectral characterization time.  The left panel of Figure \ref{revisits_impact} shows the yield as a function of aperture size with and without revisits.  For this plot, we have ignored spectral characterization time.  The dotted line shows the single visit yield while the solid line shows the yield when allowing up to 10 visits per star.

Optimized revisits increase yield by 35--75\%, with the largest changes occuring for smaller apertures.  More detailed calculations show that revisits increase yield by $\sim40\%$ across a broad range of IWAs, contrasts, and exozodi levels.  Roughly one quarter of this increase specifically comes from optimizing the delay time between revisits.  As we will show in Section \ref{sensitivity_section}, yield scales roughly as $D^{2}$.  Thus, the addition of optimized revisits reduces the required aperture by $\sim15\%$.

The right panel of Figure \ref{revisits_impact} shows the number of unique stars observed.  When including revisits, the number of stars visited decreases, as expected since stars are visited multiple times.  The number of \emph{observations} increases, though, since nearby stars that have short integration times are observed multiple times.

\subsection{The impact of spectral characterization time}

\begin{figure}[H]
\begin{center}
\includegraphics[width=6in]{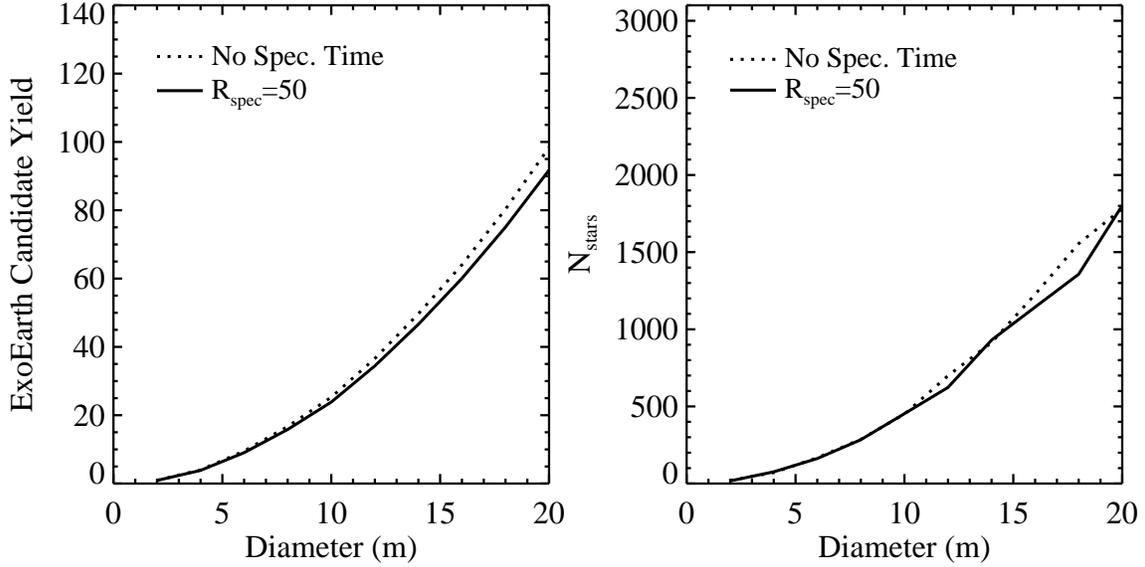}
\caption{Comparison of ExoEarth candidate yield (left) and number of unique stars observed (right) as functions of aperture size for the cases of no spectral characterization time and an $R_{\rm c}=50$, S/N$_{\rm c}=5$ spectrum obtained on every candidate exoEarth.  All calculations allow for up to 20 visits per star.  \label{sct_impact}}
\end{center}
\end{figure}

Figure \ref{sct_impact} shows how the multi-visit exoEarth candidate yield changes when we ``turn on" both revisits and our calculation of spectral characterization time.  For this plot, we use the baseline mission parameters, for which we assume noiseless detectors.  The spectral characterization time reduces yield by 3--7\%.

\subsection{Sensitivity of exoEarth candidate yield to mission parameters}
\label{sensitivity_section}

Here we vary one parameter at a time to show how the yield responds to deviations from the baseline mission.  For these results, we include optimized revisits and spectral characterization time using the parameters listed in Table \ref{baseline_params_table}.

\begin{figure}[H]
\begin{center}
\includegraphics[width=4in]{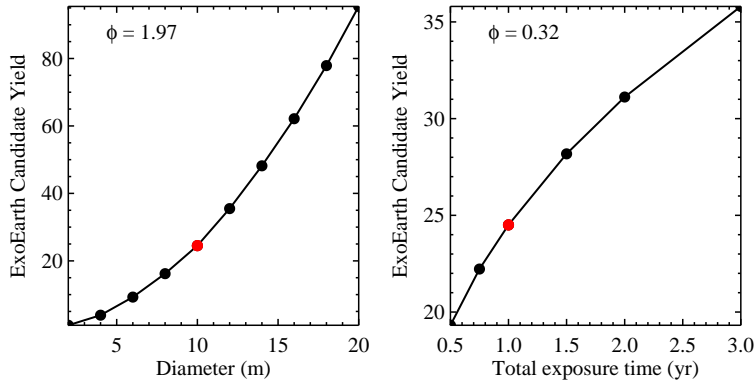}
\caption{ExoEarth candidate yield as a function of mission parameters.  Yield is a strong function of aperture and a modest function of total exposure time/mission lifetime.  \label{yield_vs_D_and_t}}
\end{center}
\end{figure}

Figure \ref{yield_vs_D_and_t} shows how the exoEarth candidate yield scales with aperture as well as total exposure time for the mission.  The red dot represents the baseline mission parameters.  Each plot is inset with a value for $\phi$, which is the approximate sensitivity of the yield with respect to the parameter being varied, i.e. the exponent of the power law relationship between yield and the parameter plotted, at the location of the red dot.

As shown on the left, near the baseline mission parameters, yield scales roughly as $D^{2}$; the mission yield can be doubled by increasing the aperture size by $\sim40\%$.  This is the strongest dependence of any parameter that we investigated, even if one considers yield to be a function of $D^2$, which would reduce $\phi$ by a factor of two.  This strong dependence is due to the fact that an increase in $D$ produces three benefits: an increase in photon collection rate, a decrease in PSF size (reducing the amount of background noise blended with the planet's signal), and a decrease in the on-sky IWA obtained with the coronagraph.  Each of these three benefits contribute roughly equally to the dependence of yield on $D$.

As shown in the right panel of Figure \ref{yield_vs_D_and_t}, the yield is a modest function of total exposure time (i.e., mission lifetime), with yield scaling as $(\Sigma\tau)^{0.32}$.  \emph{A doubling of mission lifetime only increases yield by $\sim25\%$.}  This modest dependence is a result of assuming targets can be observed in priority rank order.  As the mission progresses, one must observe lower priority targets.  Equivalently, if the mission must sacrifice total exposure time, the lowest priority targets are removed.  This selection effect, discussed in \citet{stark2014_2}, leads to moderately weak dependencies for many of the mission parameters that follow.

\begin{figure}[H]
\begin{center}
\includegraphics[width=6.5in]{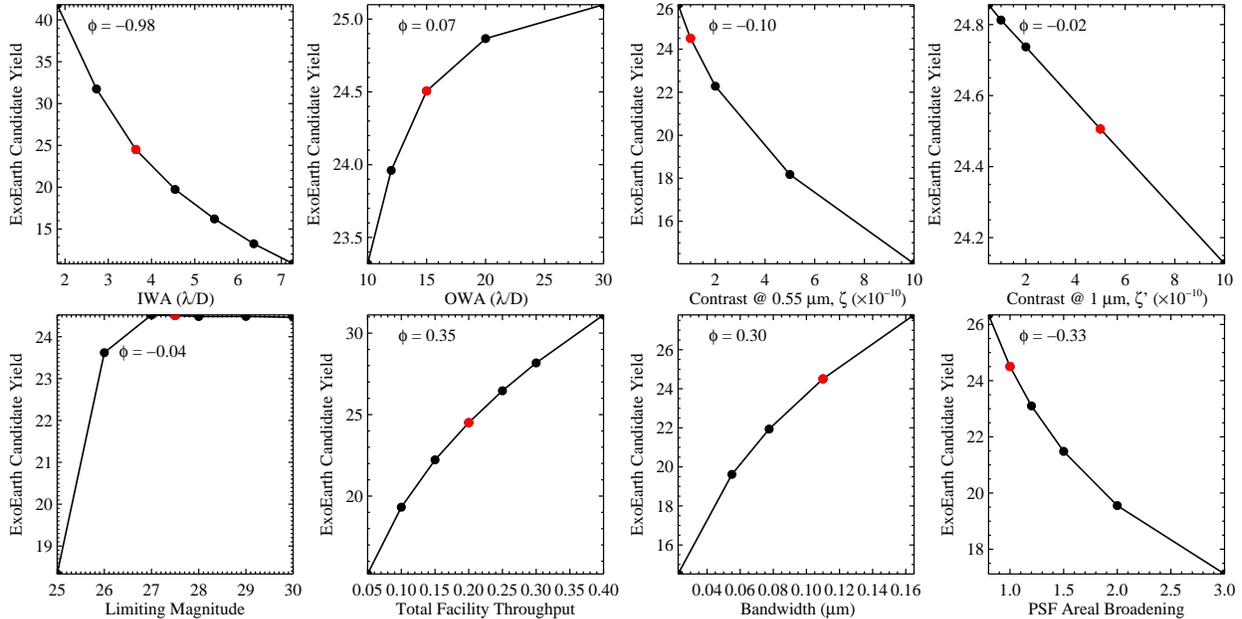}
\caption{ExoEarth candidate yield as a function of instrument parameters.  \label{yield_vs_instrument_params}}
\end{center}
\end{figure}

Figure \ref{yield_vs_instrument_params} shows the yield as a function of mission parameters that are most relevant to the internal coronagraph instrument.  The top-left panels show the yield as a function of the coronagraphs' IWA and OWA.  We remind the reader that we choose an IWA$_{\rm d}=3.6\lambda_{\rm d}/D$ for the detection coronagraph at $\lambda_{\rm d}=0.55$ $\mu$m, and an IWA$_{\rm c}=2\lambda_{\rm c}/D$ for the characterization coronagraph at $\lambda_{\rm c}=1$ $\mu$m, such that both coronagraphs have the same physical IWA.  Compared to other instrument parameters, the yield depends quite strongly on the coronagraphs' IWA near the baseline mission parameters, with yield $\propto {\rm IWA}^{-0.98}$.  An increase in the coronagraph IWA from the assumed $3.6\lambda_{\rm d}/D$ ($2\lambda_{\rm c}/D$) to $5.5\lambda_{\rm d}/D$ ($3\lambda_{\rm c}/D$) would reduce yield by $\sim35\%$.

The top-right panels in Figure \ref{yield_vs_instrument_params} show the yield as a function of the coronagraphs' contrasts, which are independent quantities.  Yield is a weak function of the detection coronagraph's contrast, with yield $\propto \zeta_{\rm d}^{-0.1}$.  An order of magnitude degradation in contrast from the assumed $10^{-10}$ to $10^{-9}$ reduces yield by $\sim40\%$.  This is slightly weaker than the dependence for the baseline parameters in \citet{stark2014_2}, primarily because here we assume a noise floor $2.5$ magnitudes fainter than the raw contrast, such that Earth twins are observable even at contrasts of $\zeta_{\rm d}=10^{-9}$.  In other words, the dependence of yield on contrast is weak in part because we assume the unsuppressed starlight contributes noise that can be overcome with longer integration times.  Contrast also only impacts nearby targets; the background count rates are dominated by local zodi and exozodi for targets farther than $\sim7$ pc, where most planets are found.

The yield is even less sensitive to the characterization coronagraph's contrast.  This is because we only require spectral characterization for those planets that are truly exoEarth candidates, i.e., exoEarth candidates can be distinguished from other types of planets by brightness, orientation, and color information.  If this assumption is invalid and spectral characterization must be performed for all observations, we expect the characterization coronagraph's contrast to be at least as important as the detection coronagraph's contrast.

The bottom-left panel in Figure \ref{yield_vs_instrument_params} shows the yield as a function of the systematic noise floor.  The systematic noise floor is plotted as the ``limiting magnitude," or the difference in magnitudes between the star and the dimmest planet that can be detected at the desired threshold S/N.  As shown in \citet{stark2014_2}, if one's goal is to find Earth twins, there is no benefit to observing planets fainter than $26.5$ magnitudes dimmer than their host star.

The bottom-middle panels in Figure \ref{yield_vs_instrument_params} show the yield as a function of throughput and bandwidth.  These two quantities are mathematically equivalent in the detection exposure time equation, so we may expect them to have identical sensitivities.  However, under our current assumptions, the throughput factors into both detection time and characterization time, whereas the bandwidth only factors into detection time.  The dependence of yield on throughput is roughly equal to that of total exposure time, as expected since they both control the number of photons collected within the mission lifetime.  The priority rank selection effect previously discussed produces the modest dependence on both of these parameters.  We note that in reality, bandwidth will affect the wavefront control overheads (which we do not model in this paper) and thus indirectly affect the spectral characterization time.

Finally, the lower-right panel shows the yield as a function of  the planet's PSF areal broadening factor, $b$.  Again, we find the same moderately weak dependence: yield is $\propto (1/b)^{0.33}$.  As shown by Equations 7--15 in \citet{stark2014_2}, in the background-limited regime in which the planet's count rate is much less than the background count rate, exposure time is inversely proportional to both throughput and $(1/\Omega)$.  A doubling of the planet's PSF size will roughly reduce the yield by $\sim25\%$.

\begin{figure}[H]
\begin{center}
\includegraphics[width=4in]{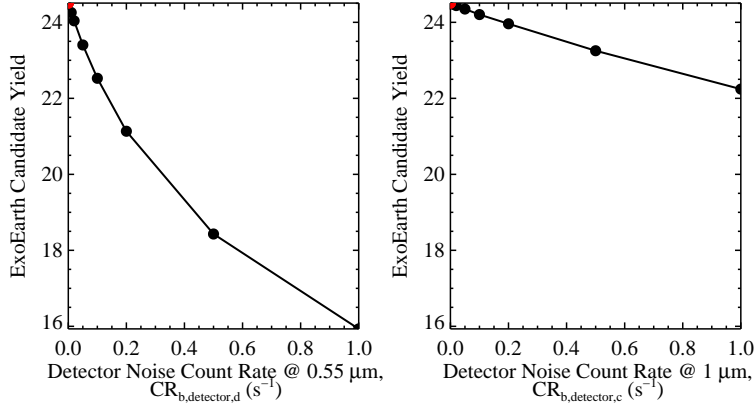}
\caption{ExoEarth candidate yield as a function of detector noise count rates. \label{yield_vs_CRbdetector}}
\end{center}
\end{figure}

Figure \ref{yield_vs_CRbdetector} shows the yield as a function of the detector noise count rates.  We have not included $\phi$ values in these plots as we assumed noiseless detectors for the baseline mission.  For the detection coronagraph, detector noise count rates $\gtrsim 0.06$ counts sec$^{-1}$ decrease the yield by $>5\%$.  For the characterization coronagraph, detector noise count rates $\gtrsim 0.5$ counts sec$^{-1}$ decrease the yield by $>5\%$.  Clearly the yield depends more strongly on the noise count rate of the detection coronagraph.  However, this does not necessarily mean that the required read noise and dark current of the detection coronagraph's detector are lower.

To translate the above count rates into combinations of read noise and dark current (RN, $\xi$), we use Equation \ref{detector_noise_eq} and assume values for $\tau_{\rm expose}$ and $n_{\rm pix}$.   We assume the length of an individual exposure $\tau_{\rm expose} = 1000$ s to allow for the removal of cosmic ray artifacts.  The number of pixels over which the planet's photometric aperture is spread, $n_{\rm pix}$, depends on the optical layout of the instrument.

For detection, we imagine three types of instruments that could be used: a simple imager (assuming spectral information is not needed for wavefront control), an imager with a frequency-resolving detector, and an integral field spectrograph (IFS).  If the detection coronagraph is a simple imager or is frequency-resolving, the planet's photometric aperture would be spread over $\sim4$ pixels assuming imperfect centering and Nyquist sampling of the PSF.  If the detection coronagraph is an IFS, then the planet's photometric aperture would be spread over 4 lenslets, each of which is dispersed onto 6 pixels (2 in the frequency domain and 3 in the spatial domain) per spectral channel.  Assuming $R=50$, such that there are 10 spectral channels over the 20\% detection bandpass, the IFS would spread the planet's photometric aperture over 240 pixels.

For spectral characterization, we imagine two types of instruments: an imager with energy-resolving detector and an IFS.  Similar to a broadband detection, the planet's photometric aperture would be spread over 4 pixels for each spectral channel in an energy-resolving detector.  As discussed previously, the IFS would spread the planet's photometric aperture over 24 pixels per spectral channel (4 lenslets, 6 pixels per lenslet).

\begin{figure}[H]
\begin{center}
\includegraphics[width=6.5in]{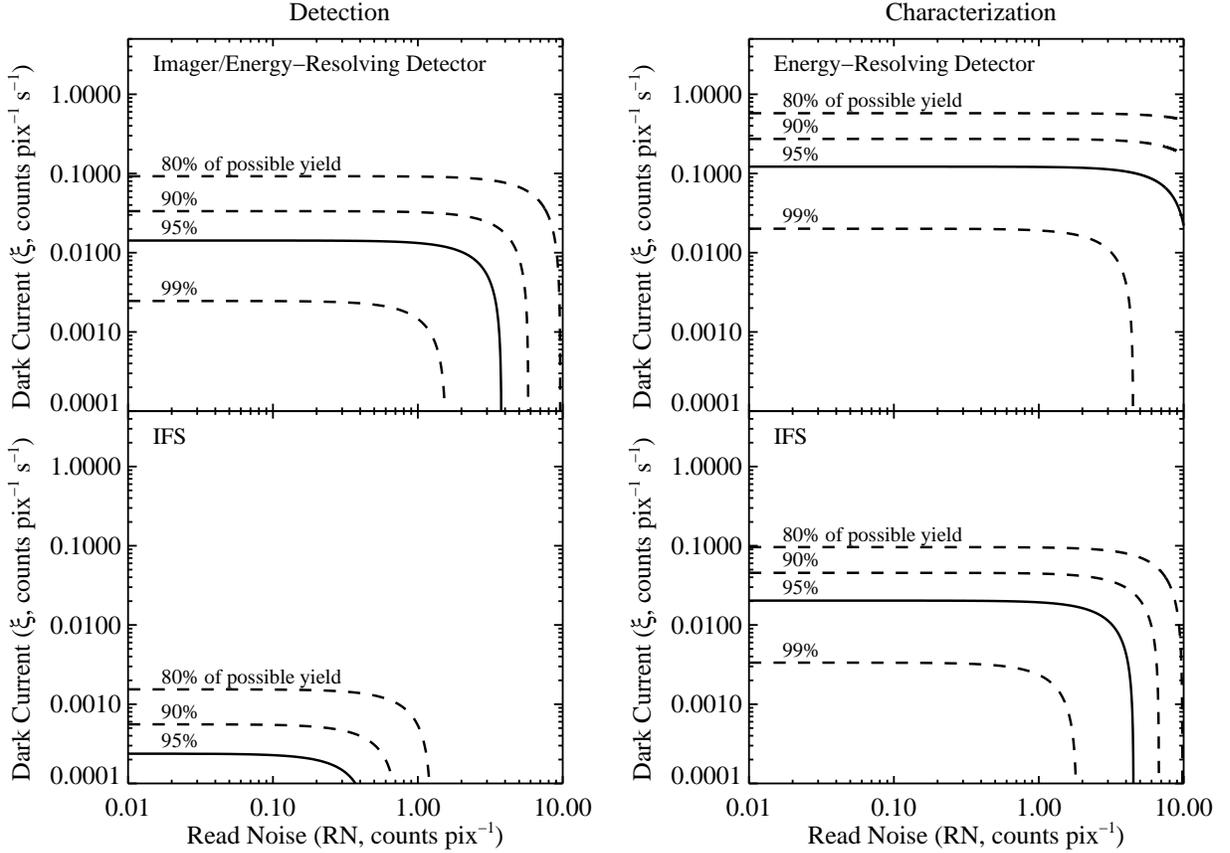}
\caption{ExoEarth candidate yield as a function of read noise and dark current, normalized to the noiseless detector case, for different detection coronagraph instruments (left) and characterization coronagraph instruments (right).  Combinations of read noise and dark current to the lower-left of the solid line are nominally acceptable for the baseline mission parameters.  All curves are sensitive to bandwidth/spectral resolving power and the required signal to noise ratio (see Figure \ref{detector_noise_vs_r_vs_snr}). \label{detector_noise_curves}}
\end{center}
\end{figure}

Figure \ref{detector_noise_curves} shows the fraction of the yield realized for different instruments.  The left panels show the impact of detector noise on yield for the V band detection coronagraph only.  The right panels show the impact of detector noise on yield for the 1 $\mu$m characterization coronagraph only.  These curves were created by interpolating the yield curves from Figure \ref{yield_vs_CRbdetector} and adopting the values of $n_{\rm pix}$ described above.   They were calculated without any exposure time limitations due to stability requirements or motion of the observed planet.

\begin{figure}[H]
\begin{center}
\includegraphics[width=4in]{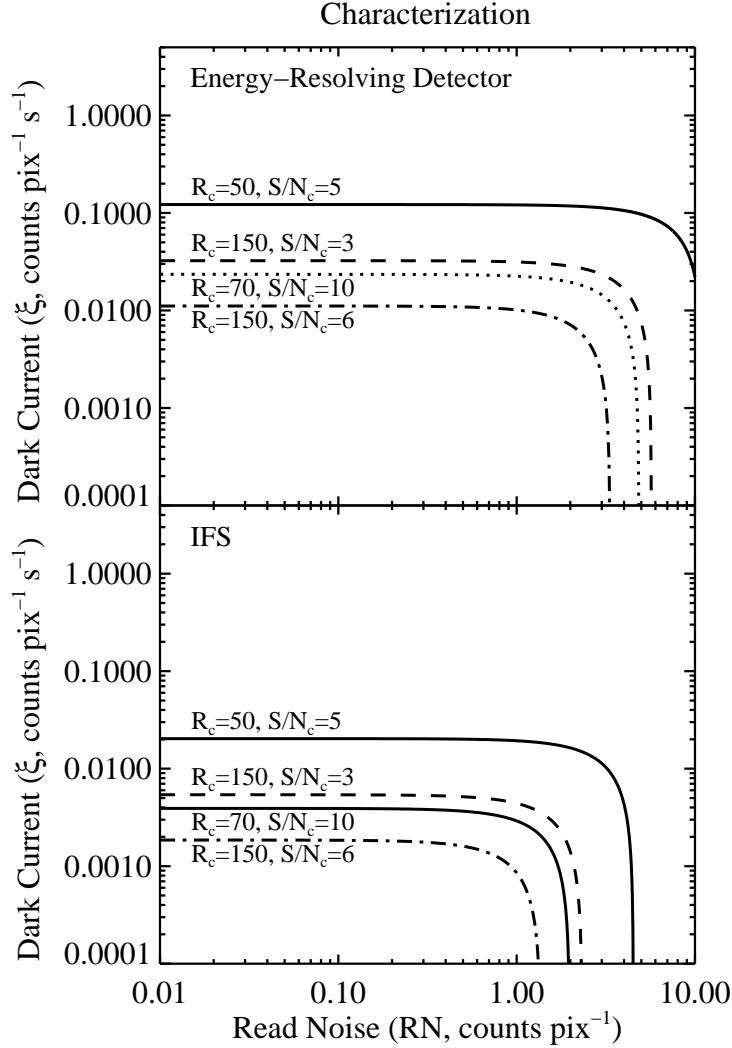}
\caption{Curves for 95\% of possible yield (see Figure \ref{detector_noise_curves}) for the characterization coronagraph's detector for different combinations of $R_{\rm c}$ and S/N$_{\rm c}$.  Estimates of the allowable detector noise parameters are sensitive to bandwidth/spectral resolving power and S/N. \label{detector_noise_vs_r_vs_snr}}
\end{center}
\end{figure}

While these (RN, $\xi$) curves are roughly independent of aperture size, they are sensitive to bandwidth/spectral resolving power and S/N.  Figure \ref{detector_noise_vs_r_vs_snr} shows the $95\%$ yield curves for the characterization coronagraph's detector for several combinations of $R_{\rm c}$ and S/N$_{\rm c}$.  We discuss the implications of these curves in Section \ref{discussion}.

\begin{figure}[H]
\begin{center}
\includegraphics[width=3in]{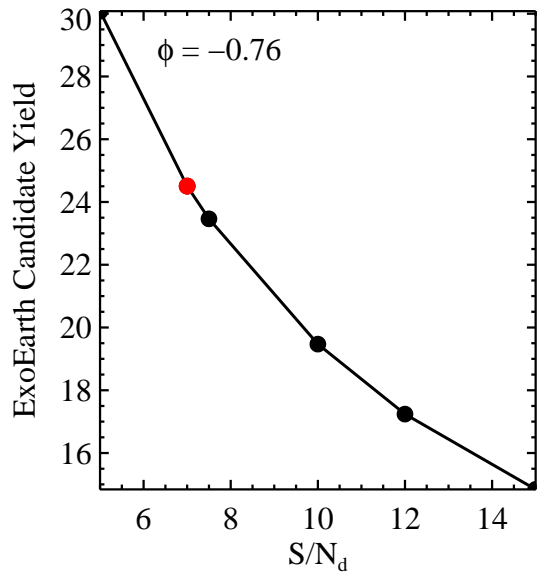}
\caption{ExoEarth candidate yield as a function of threshold S/N required for detection.  Yield strongly depends on S/N$_{\rm d}$. \label{yield_vs_SNR}}
\end{center}
\end{figure}

Figure \ref{yield_vs_SNR} shows the yield as a function of the threshold detection S/N, which shows a strong dependence.  The threshold S/N$_{\rm d}$ is not a telescope or instrument parameter---it is a numerical parameter that defines the minimum signal to noise ratio allowable for detection.  In our calculations, we assume that the signal to noise ratio of an exoEarth candidate follows Poisson statistics; we estimate the noise by simply taking the square root of the number of counts.  If this were truly the case, we could justifiably choose a threshold S/N$_{\rm d}$ as low as 5.  In reality, unknown systematic uncertainties and complexities with background modeling/subtraction will likely reduce the true S/N, motivating previous works to choose a threshold S/N$_{\rm d}=10$ \citep[e.g.,][]{brown2005,turnbull2012,stark2014_2}.

\begin{figure}[H]
\begin{center}
\includegraphics[width=3in]{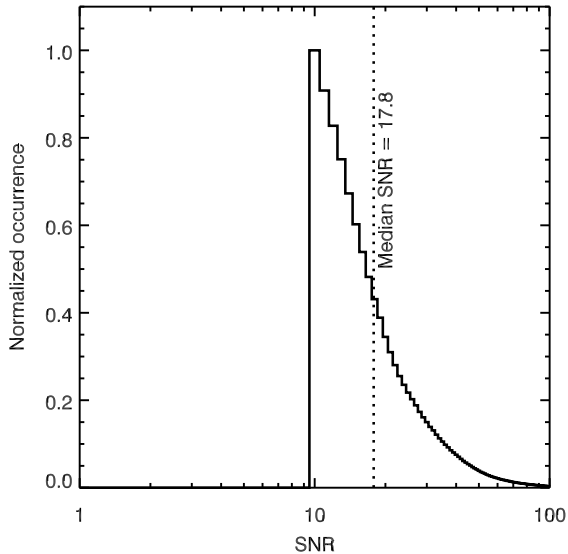}
\caption{Realized S/N for all synthetic exoEarths detectable with the baseline mission assuming a threshold S/N$_{\rm d}=10$.  A threshold S/N$_{\rm d}=10$ is too conservative, leading to a median S/N $=18$.  \label{SNR_histo_fig}}
\end{center}
\end{figure}

Figure \ref{SNR_histo_fig} shows a histogram of realized S/N for the baseline mission when adopting a threshold S/N$_{\rm d}=10$.  Because our Monte Carlo calculations produce synthetic exoEarths covering a broad range of phase angles and circumstellar distances, the ``cloud" of exoEarths around a given star has a broad distribution of planet brightness.  Our code tunes the exposure time to the dimmest planet desired for detection using the threshold S/N$_{\rm d}$.  As a result, all other potentially detectable planets around a star, which are by definition brighter, will have realized S/N $>$ S/N$_{\rm d}$ for the chosen exposure time.

As shown by Figure \ref{SNR_histo_fig}, a threshold S/N$_{\rm d}=10$ produces a median S/N $=18$.  This ``typical" S/N is far more conservative than necessary.  In light of this, we have reduced the baseline threshold S/N$_{\rm d}$ for this study.  We adopt a baseline threshold S/N$_{\rm d}=7$, which produces a median S/N $=12.5$, motivated by the fact that S/N$_{\rm d}=7$ provides a low false-negative rate \citep{kasdin2006}.

\subsection{Sensitivity of exoEarth candidate yield to astrophysical parameters}

\begin{figure}[H]
\begin{center}
\includegraphics[width=6in]{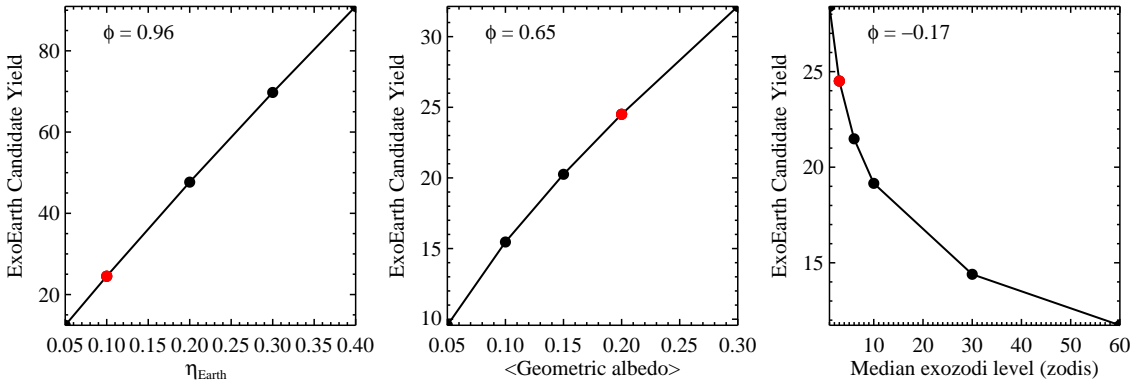}
\caption{ExoEarth candidate yield as functions of astrophysical parameters.  \label{yield_vs_astro_fig}}
\end{center}
\end{figure}

Figure \ref{yield_vs_astro_fig} shows how the exoEarth yield responds to changes in astrophysical parameters.  The left panel shows that yield is a strong function of $\eta_{\earth}$, with $\phi = 0.96$.  The yield is not quite linearly proportional to $\eta_{\earth}$, because larger values of $\eta_{\earth}$ demand more characterization time at the expense of detection time.  Other combinations of mission parameters reveal that the dependence on $\eta_{\earth}$ can be even weaker, with $\phi \lesssim 0.8$.

The middle panel of Figure \ref{yield_vs_astro_fig} shows that yield also strongly depends on the typical geometric albedo of the exoEarth candidates, a parameter we likely cannot constrain prior to a direct imaging mission.  Our baseline mission assumes a geometric albedo for all exoEarth candidates equal to that of the Earth.  If the typical exoEarth candidate albedo is less than that of Earth (e.g., because many of the planets do not have cloud cover), we will miss some of the low-albedo exoEarth candidates, but the probability of detecting the Earth-twins remains the same.

Finally, the panel on the right shows the sensitivity of the yield to changes in the median exozodi level.  Near the baseline mission parameters $\phi=-0.17$, such that a factor of 10 increase in the median exozodi level reduces yield by $\sim35\%$.  As explained by \citet{stark2014_2}, this weak dependence is due to the priority selection effect previously discussed, as well as changes to the optimum observation strategy: for larger exozodi levels, planet searches should limit themselves to more gibbous phases.

We note that although the right-most panel in Figure \ref{yield_vs_astro_fig} is labeled ``median exozodi level," we in fact set every target to the same exozodi level.  Of course the universe will deal a distribution of exozodi levels, so it is important for us to consider the impact of such a distribution.  To investigate this, we assigned each star in our target list a random exozodi level drawn from a distribution.  We considered two distributions: a uniform distribution from 0 to $2n_{\rm median}$, and a log-normal distribution with a median exozodi level $n_{\rm median}$ and $\sigma$ parameter of $1.25$ \citep{mennesson2014}.  We ran 15 independent instances of the yield maximization code, randomizing the exozodi levels each time.

\begin{figure}[ht]
\centering
\begin{minipage}[b]{0.45\linewidth}
\centering
\includegraphics[width=.99\linewidth]{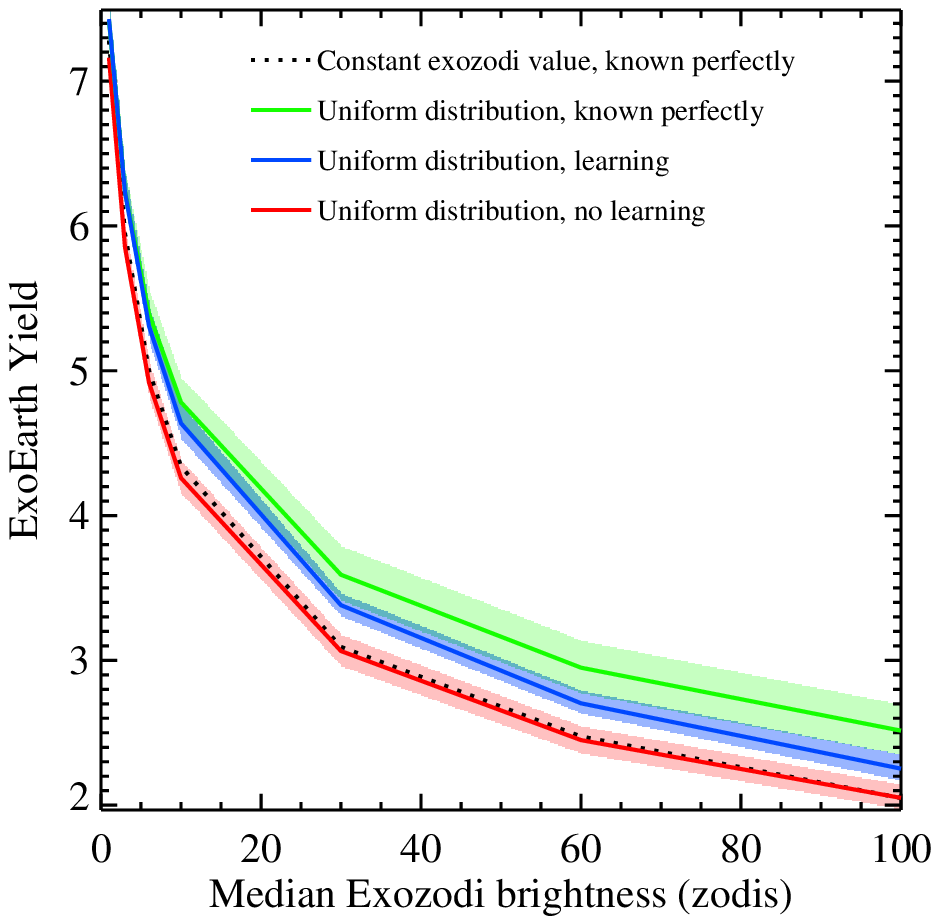}
\label{fig:minipage1}
\end{minipage}
\quad
\begin{minipage}[b]{0.45\linewidth}
\centering
\includegraphics[width=.99\linewidth]{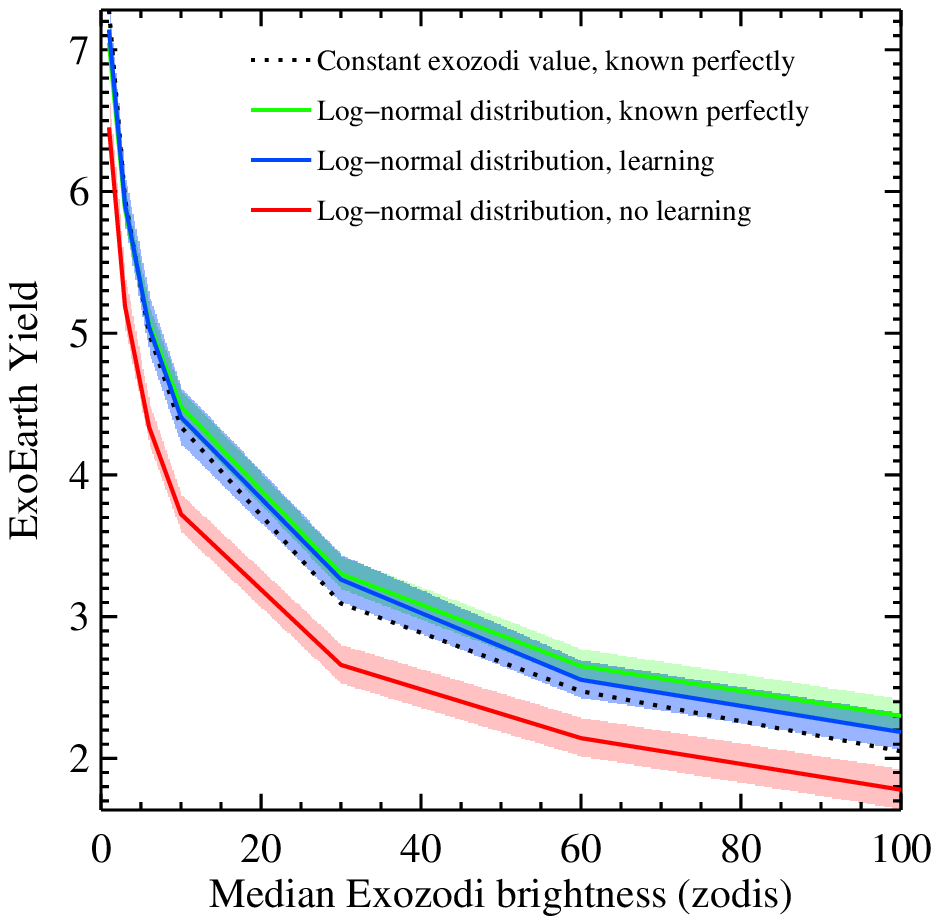}
\label{fig:minipage2}
\end{minipage}
\caption{ExoEarth candidate yield as a function of median exozodi level for a uniform distribution of exozodi levels (left) and a log-normal distribution (right).  Updating the observation plan as individual exozodi levels are measured can mitigate any negative impacts of the unknown exozodi distribution. \label{yield_vs_exozodi_dist}}
\end{figure}

Figure \ref{yield_vs_exozodi_dist} shows the yield as a function of exozodi level for a uniform distribution (left) and a log-normal distribution (right).  For this figure, we adopted a 4 m telescope aperture and ignored spectral characterization time to reduce the run time of the code.  The black dotted line in each plot represents the yield if all stars have the same exozodi level.  The green curves (and green shaded regions) represent the yield (and $1\sigma$ deviation) if every star's individual exozodi level is known ahead of time.

The red curve shows the yield if we know the median in advance, but no individual levels, and refuse to adjust the observation plan as the mission progresses.  Assuming we know the median exozodi level in advance is a valid approximation because we should be able to estimate the median after a dozen or so observations.  However, refusing to adjust the observation plan as we learn individual exozodi levels makes this a conservative estimate.

The blue curve shows the yield if we know the median exozodi level in advance (or very early in the mission), and we update the observation plan as we observe each star and measure its exozodi level.  To calculate this curve, we first ran the yield maximization code under the assumption that all stars have an exozodi level equal to the median exozodi level to obtain an initial observation plan.  Then we ran the code again, allowing one star's true exozodi level to be known.  We locked that star's observations into the plan, prohibiting them from being adjusted later, then ran the code again with one more star's true exozodi level known.  As each star was observed and its exozodi level was measured, we allowed the code to re-optimize all unobserved stars.  We assumed stars were observed in priority order and continued this process until the baseline total exposure time was reached.

As shown in Figure \ref{yield_vs_exozodi_dist}, a distribution of exozodi level does not dramatically impact the mission yield.  Our approximation of all stars having an exozodi level equal to the median level is valid.  As long as we allow the observation plan to adjust to individual star's exozodi levels, we can avoid the negative impacts of the unknown exozodi distribution and, instead, obtain small increases in the exoEarth candidate yield.

\section{Discussion}
\label{discussion}

The scaling relationships derived in this work are very similar to those derived in \citet{stark2014_2}.  Surprisingly, while revisits and spectral characterization time affect the absolute yield, they do not significantly impact the scaling relationships.

The sensitivity parameters ($\phi$ values) calculated in the previous section inform us of the approximate power law relationships between yield and each parameter when deviating from the baseline mission.  However, they do not necessarily tell us which mission parameters will ultimately produce the greatest increase in yield.  For example, the coronagraph IWA sensitivity ($\phi=-0.98$) is significantly larger than the detection coronagraph's contrast sensitivity ($\phi=-0.10$).  An improvement of the coronagraph IWA by just $0.5\lambda_{\rm c}/D$ would increase yield by $\sim30\%$.  To produce an equivalent yield increase via the detection coronagraph's contrast would require roughly a factor of 10 improvement.  But given current coronagraph contrasts and IWAs, both of these may be equally likely/challenging \citep{shaklan2013}.

Similarly, although the yield is weakly dependent on exozodi level ($\phi=-0.17$), we have relatively poor constraints on the median exozodi level.  The current upper-limit on median exozodi level is 60 zodis \citep{mennesson2014}.  In the absence of future constraints, we should plan a mission capable of dealing with this level of exozodiacal dust to avoid unnecessary risk.  If we can constrain the median exozodi level to $\sim3$ zodis, which may be possible with the Large Binocular Telescope Interferometer (LBTI), the yield would increase by more than a factor of two, reducing the required aperture by $\sim30\%$.

The yield strongly depends on the threshold S/N$_{\rm d}$, i.e., the definition of a ``detection."  In Section \ref{results_section}, we justified relaxing this parameter from the commonly adopted value of 10 to our baseline assumption of 7.  This alone increased yield by $30\%$.  Additional reductions may be justifiable as well; reducing the threshold S/N$_{\rm d}$ to 5 would provide another $30\%$ increase in yield.  However, we feel that more detailed simulations of telescope/instrument performance and exoEarth signal extraction are required to justify any further reductions.

We currently assume one year of total exposure time out of an assumed five year mission.  A coronagraph could require one additional year of overheads to achieve the desired contrast on every target.  Thus, our assumptions could implicitly require two years of a five year mission.  Given that yield weakly depends on total exposure time ($\phi = 0.33$), increases to the total exposure time provide little additional yield and would require a larger fraction of the total mission.  We therefore find it unlikely that increasing the time devoted to exoEarth science for a general astrophysics missions could increase yield by more than $\sim10\%$.

The weak total exposure time dependence also argues against a multi-telescope coronagraph mission, because flying $N$ telescopes is equivalent to one telescope lasting $N$ times as long.  For example, to double the exoEarth candidate yield, one can either increase the aperture by $45\%$ or fly \emph{eight} identical telescopes.  Cost scalings likely make the former a more attractive option.

Under the baseline assumptions of zero detector noise, spectral characterization does not strongly impact the yield.  This assumes that $R_{\rm c}=50$ spectra are obtained only for the true exoEarth candidates and equivalent spectra are not obtained for potentially confusing crescent phase gas giants or other background objects.  Depending on the observing strategy implemented, this may be a reasonable approximation.  However, the weak impact of spectral characterization time is also a result of the weak dependence of yield on total exposure time.  Even if \emph{half} of the total exposure time is devoted to spectral characterization, the mission yield is only reduced by $20\%$.  

If detector noise is non-zero, the increase in the required exposure times can significantly impact yield.  To define acceptable detector noise parameters, we must choose an acceptable level of yield reduction from Figure \ref{detector_noise_curves}.  We choose a $10\%$ reduction in yield as a maximally acceptable value, which could be offset by a $5\%$ increase in aperture size.  The impact of noise from the detection and characterization coronagraphs' detectors will compound, so we allow for at most a $5\%$ reduction in yield due to each of the coronagraphs' detectors.  These ``allowable" yield reductions are shown as solid lines in Figure \ref{detector_noise_curves}.  Any noise properties to the lower-left of these curves would be acceptable from a yield perspective for our baseline mission parameters.  

Because an IFS spreads the planet's light over a much larger number of pixels, tolerable noise properties are much more stringent.  For planet detection with an IFS under the baseline assumptions, we require a detector with dark current $\lesssim 0.0001$ pix$^{-1}$ s$^{-1}$ and read noise $\lesssim 0.2$ pix$^{-1}$.  Energy-resolving detectors, which spread the planet's light over a small number of pixels, must only achieve dark currents $\lesssim 0.01$ pix$^{-1}$ s$^{-1}$ and read noise $\lesssim 2$ pix$^{-1}$, on par with current CCD noise performance.

We note, however, that we have not considered exposure time limitations due to pointing requirements or the motion of the planet being observed.  The allowable noise parameters for the characterization coronagraph's detector may become more constrained as one limits the spectral characterization time.  Further, the allowable noise properties of the characterization coronagraph's detector are sensitive to the values of $R_{\rm c}$ and S/N$_{\rm c}$, as shown by Figure \ref{detector_noise_vs_r_vs_snr}.  Thus the tolerable detector noise properties should be defined in terms of the most difficult spectral line one wants to detect.  For our baseline mission in this work, we considered only the detection of water near 1 $\mu$m, which requires $R_{\rm c}\sim50$ and S/N$_{\rm c}\sim5$.  If one assumes the mission must be capable of detecting O$_2$ near $0.75$ $\mu$m, such that $R_{\rm c}\sim150$ and S/N$_{\rm c}\sim6$ \citep{brandt2014}, then the dot-dashed curves in Figure \ref{detector_noise_vs_r_vs_snr} should be considered in defining the tolerable noise parameters.

\subsection{Lower limits on aperture size}

For our baseline mission, we assume noiseless detectors, relatively high total throughput of 20\% for a coronagraphic mission, raw detection contrast of $10^{-10}$, and a $360^{\circ}$ detection zone between the IWA and OWA.  Additional sources of performance degradation, like jitter, are not considered and implicitly assumed to be negligible.  We can therefore consider our baseline coronagraph performance optimistic.

Let's assume for the moment that such an optimistic mission is attainable and examine the yield as a function of telescope aperture under different astrophysical assumptions.  In a pessimstic/low risk scenario, we would design a mission robust to a low value of $\eta_{\earth}$.  Thus, for our pessimistic case we choose $\eta_{\earth} = 0.1$, our 1$\sigma$ lower limit from Table \ref{eta_Earth_table}.  It is extremely unlikely that a mission would be planned using anything greater than the expected value for $\eta_{\earth}$, so for the optimistic/high risk scenario we adopt $\eta_{\earth} = 0.16$, our expected value for the OKHZ.

\begin{figure}[H]
\begin{center}
\includegraphics[width=6.5in]{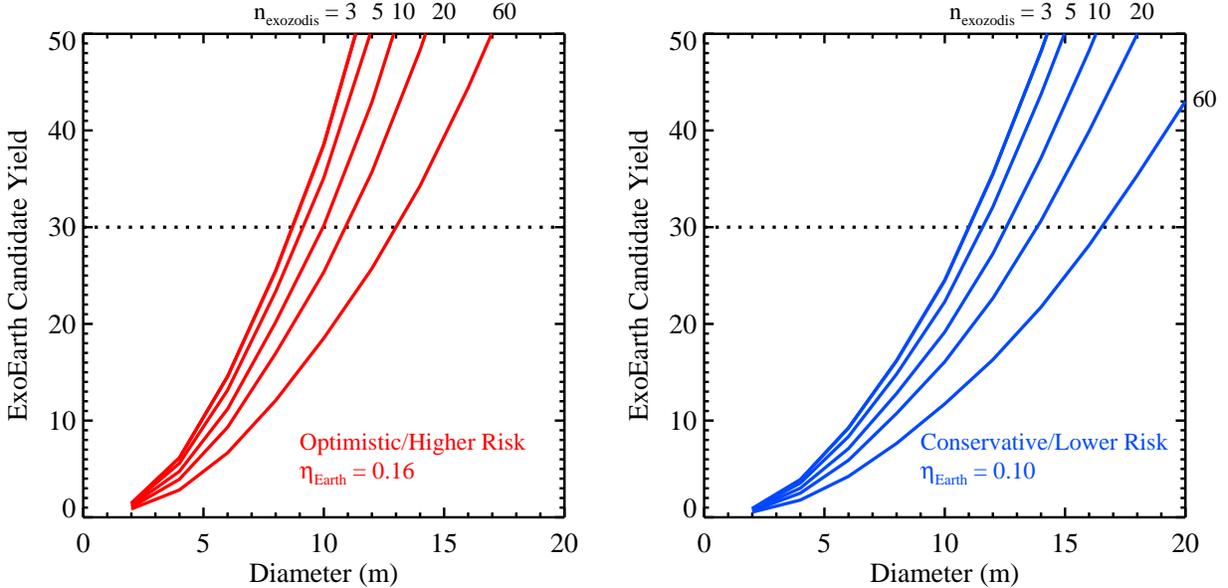}
\caption{ExoEarth candidate yield vs aperture for the optimistic (higher risk) and pessimistic (lower risk) values of $\eta_{\earth}$.  Our minimum acceptable yield for an exoEarth characterizing mission is shown as a dotted line.  To exceed the minimum yield under the most optimistic (and potentially unrealistic) assumptions, we require an aperture $>8.5$ m.  \label{minimum_aperture_fig}}
\end{center}
\end{figure}

Figure \ref{minimum_aperture_fig} shows the exoEarth candidate yield as a function of aperture size and median exozodi level under these two assumptions.  The current constraint on the median exozodi level is $\le60$ zodis \citep{mennesson2014}, the rightmost curve in each plot.  If LBTI can further constrain the median exozodi level, perhaps as low as $\sim$3 zodis, the yield would follow the curves on the left.

To place a lower limit on the required aperture size, we must choose a minimum acceptable yield.  Equation 3 from \citet{stark2014_2} provides a quantitative means to define a yield goal.  Here, we present a modified form that accounts for the counting statistics of all exoEarth candidates, not just the ``true" exoEarths:
\begin{equation}
	\label{yield_goal_eq}
	N_{\rm EC} = \eta_{\earth}\; \frac{\log{(1 - C)}}{\log{(1 - \eta_{\earth} f_x)}},
\end{equation}
where $N_{\rm EC}$ is the target yield of exoEarth candidates, $C$ is the confidence of finding at least one candidate with $x$ detectable in its atmosphere, $f_x$ is the fraction of exoEarth candidates that have $x$ detectable in their atmospheres, and $x$ can be water, oxygen, methane, or anything else scientifically compelling.  

\citet{stark2014_2} suggested designing a mission robust to $f_x=0.1$ with $99.7\%$ confidence.  To place a lower limit on the aperture size, we relax this constraint to $f_x=0.1$ with $95\%$ confidence, which we consider the minimum acceptable yield.  Assuming that detectable levels of $x$ define an ``Earth-like" planet, designing a mission to this standard guarantees, with 95\% confidence, the detection of at least one Earth-like planet if one in ten Earth-sized planets in the HZ are Earth-like.  Substituting these values into Equation \ref{yield_goal_eq} gives a yield goal of 30 exoEarth candidates for both the optimistic and pessimistic values of $\eta_{\earth}$.

We plot the minimum scientifically compelling yield goal as a horizontal dashed line in Figure \ref{minimum_aperture_fig}.  Even in the most optimistic scenario possible, with an optimistic coronagraph, optimistic astrophysical assumptions, and an optimistic observation plan in which spectral characterization is only performed on those planets that are truly exoEarth candidates, we require an aperture $>8.5$ m to achieve the minimum yield goal.  Given that the planned Space Launch System (SLS) Block-2 with 10 m fairing, the most capable space launch rocket for the foreseeable future, could not launch a monolithic aperture $>8$ m in diameter, a coronagraphic exoEarth detecting mission may require a segmented aperture.

We note that Figure \ref{minimum_aperture_fig} assumes no change in telescope or coronagraph performance as a function of $D$.  In reality, segmented apertures require specialized coronagraphs, so we expect some change in performance at the transition from monolithic to segmented apertures, which will occur at $\le8$ m given the largest SLS fairing diameter.  To date, the majority of coronagraph design work has focused on monolithic, off-axis apertures.  Current coronagraphs designed to work with segmented apertures have comparatively lower performance, though there are designs that, in principle, can perform on-par with off-axis designs \citep{mamadou2015,lyon2015}.  Thus, if one were to plot yield as a function of aperture for today's coronagraphs, a significant reduction in yield would occur when crossing the 8 m threshold, which would eventually be offset at larger apertures.  This degradation in yield at the 8 m transition is not well quantified and depends on the parameters of the specific coronagraphs chosen for monolithic and segmented apertures.  We plan to address this issue and the yield of specific coronagraphs in future work.

Finally, we note that improving the constraint on the median exozodi level can substantially reduce the required aperture.  A mission robust to 60 zodis and $\eta_{\earth} = 0.1$ requires an aperture $>16$ m.  If LBTI finds that the median exozodi level $\lesssim 3$ zodis, we would require an aperture of $\sim11$ m.

\section{Conclusions}
\label{conclusions}

We have developed a multi-wavelength completeness calculator that handles multiple visits to every star and simultaneously optimizes the number of visits, delay time between visits, and exposure time of every observation to maximize exoEarth candidate yield for a coronagraphic direct-imaging mission.  We have updated our astrophysical assumptions based on recent literature, using circular orbits within the optimistic habitable zone as defined by \citet{kopparapu2013}.  For a population of rocky planets residing in this classical habitable zone with median radius $=1$ $R_{\earth}$, we find $\eta_{\earth} = 0.16 \pm 0.06$ by scaling the results of \citet{petigura2013_2}.

We find that optimized revisits increase yields by 35--75\%, providing a $\sim15\%$ reduction in the required aperture size.  We find scaling results similar to those of \citet{stark2014_2}.  Yield is most strongly dependent on aperture size, scaling roughly as $\propto D^{2}$.  We find a moderately weak dependence of yield on total exposure time ($\propto \Sigma\tau^{0.33}$), which argues against significant increases in exoEarth science time as well as mission concepts with multiple small telescopes.

As long as post-processing enables the reliable detection of point sources brighter than $26.5$ mags dimmer than the host star, the yield is a weak function of the coronagraph's raw contrast and the contrast floor; a factor of 10 degradation in the detection coronagraph's contrast from the baseline value of $10^{-10}$ reduces yield by $\sim40\%$.  The yield is strongly dependent on the inner working angle of the coronagraph ($\propto {\rm IWA}^{-0.98}$); a degradation of the inner working angle from the baseline value of $2\lambda_{\rm c}/D$ to $3\lambda_{\rm c}/D$ reduces yield by $\sim35\%$.  We find a weak to moderately weak dependence on all other high level coronagraph parameters: outer working angle, throughput, bandwidth, and PSF sharpness.  As long as spectra are obtained for a small subset of the observations (e.g., only for likely exoEarth candidates), the yield is a very weak function of the characterization coronagraph's contrast.

Yield is a weak function of the median exozodi level, but there is considerable room for improvement in median exozodi level constraints.  Constraining the median exozodi level to $<3$ zodis, which may be possible with LBTI, could more than double the expected yield and reduce the required aperture size by 40\%.  We find little benefit in knowing the exozodi levels of individual stars a priori and find that the precise distribution of exozodi negligibly impacts yield for apertures $\gtrsim4$ m.

We have included the effects of detector noise and spectral characterization time and calculated the yield under the assumption that imaging and spectral characterization are drawn from the same total time budget.  In the limit of noiseless detectors, the required spectral characterization time does not significantly impact mission yield.  

If spectral information is not required for wavefront control and a simple broadband imager can be used for detection, or if an energy-resolving detector can be used, constraints on allowable detector noise are modest, with dark current $\lesssim 0.01$ s$^{-1}$ pixel$^{-1}$ and read noise $\lesssim 2$ pixel$^{-1}$.  If an energy-resolving detector is not an option and an IFS is required for detection, we require improved detector noise performance, with dark current $\lesssim 0.0001$ s$^{-1}$ pixel$^{-1}$ and read noise $\lesssim 0.2$ pixel$^{-1}$.  

We place lower limits on the aperture size for a number of astrophysical assumptions, while assuming optimistic telescope and coronagraph parameters.  We conclude that a coronagraphic direct-imaging mission requires a $>8.5$ m aperture to provide a reasonable chance of finding an Earth-like planet.

Future yield estimates could benefit from an improved target list, vetted in more detail for completeness, binarity, and stellar diameter.  A more realistic simulation of coronagraph performance, including contrast as a function of separation and Airy throughput as a function of separation, would enable us to compare different coronagraph designs on an equal footing.  Finally, future efforts should also focus on analogous yield maximization methods optimized for starshade-based mission concepts. 

\acknowledgments

This research was supported by an appointment to the NASA Postdoctoral Program at Goddard Space Flight Center, administered by Oak Ridge Associated Universities through a contract with NASA.  S.D.D.G., A.M., and A.R. acknowledge support by GSFC's internal research and development fund.

\bibliography{ms_v2.bbl}

\begin{thebibliography}{26}
\expandafter\ifx\csname natexlab\endcsname\relax\def\natexlab#1{#1}\fi

\bibitem[{{Brandt} \& {Spiegel}(2014)}]{brandt2014}
{Brandt}, T.~D., \& {Spiegel}, D.~S. 2014, Proceedings of the National Academy
  of Science, 111, 13278

\bibitem[{{Brown}(2005)}]{brown2005}
{Brown}, R.~A. 2005, \apj, 624, 1010

\bibitem[{{Brown} \& {Soummer}(2010)}]{brownsoummer2010}
{Brown}, R.~A., \& {Soummer}, R. 2010, \apj, 715, 122

\bibitem[{{Cumming} {et~al.}(2008){Cumming}, {Butler}, {Marcy}, {Vogt},
  {Wright}, \& {Fischer}}]{cumming2008}
{Cumming}, A., {Butler}, R.~P., {Marcy}, G.~W., {et~al.} 2008, \pasp, 120, 531

\bibitem[{{Dong} \& {Zhu}(2013)}]{dong2013}
{Dong}, S., \& {Zhu}, Z. 2013, \apj, 778, 53

\bibitem[{{Foreman-Mackey} {et~al.}(2014){Foreman-Mackey}, {Hogg}, \&
  {Morton}}]{foremanmackey2014}
{Foreman-Mackey}, D., {Hogg}, D.~W., \& {Morton}, T.~D. 2014, \apj, 795, 64

\bibitem[{{Howard} {et~al.}(2012){Howard}, {Marcy}, {Bryson}, {Jenkins},
  {Rowe}, {Batalha}, {Borucki}, {Koch}, {Dunham}, {Gautier}, {Van Cleve},
  {Cochran}, {Latham}, {Lissauer}, {Torres}, {Brown}, {Gilliland}, {Buchhave},
  {Caldwell}, {Christensen-Dalsgaard}, {Ciardi}, {Fressin}, {Haas}, {Howell},
  {Kjeldsen}, {Seager}, {Rogers}, {Sasselov}, {Steffen}, {Basri},
  {Charbonneau}, {Christiansen}, {Clarke}, {Dupree}, {Fabrycky}, {Fischer},
  {Ford}, {Fortney}, {Tarter}, {Girouard}, {Holman}, {Johnson}, {Klaus},
  {Machalek}, {Moorhead}, {Morehead}, {Ragozzine}, {Tenenbaum}, {Twicken},
  {Quinn}, {Isaacson}, {Shporer}, {Lucas}, {Walkowicz}, {Welsh}, {Boss},
  {Devore}, {Gould}, {Smith}, {Morris}, {Prsa}, {Morton}, {Still}, {Thompson},
  {Mullally}, {Endl}, \& {MacQueen}}]{howard2012}
{Howard}, A.~W., {Marcy}, G.~W., {Bryson}, S.~T., {et~al.} 2012, \apjs, 201, 15

\bibitem[{{Hunyadi} {et~al.}(2007){Hunyadi}, {Shaklan}, \&
  {Brown}}]{hunyadi2007}
{Hunyadi}, S.~L., {Shaklan}, S.~B., \& {Brown}, R.~A. 2007, in Society of
  Photo-Optical Instrumentation Engineers (SPIE) Conference Series, Vol. 6693,
  Society of Photo-Optical Instrumentation Engineers (SPIE) Conference Series

\bibitem[{{Kane} {et~al.}(2012){Kane}, {Ciardi}, {Gelino}, \& {von
  Braun}}]{kane2012}
{Kane}, S.~R., {Ciardi}, D.~R., {Gelino}, D.~M., \& {von Braun}, K. 2012,
  \mnras, 425, 757

\bibitem[{{Kasdin} \& {Braems}(2006)}]{kasdin2006}
{Kasdin}, N.~J., \& {Braems}, I. 2006, \apj, 646, 1260

\bibitem[{{Kopparapu} {et~al.}(2013){Kopparapu}, {Ramirez}, {Kasting}, {Eymet},
  {Robinson}, {Mahadevan}, {Terrien}, {Domagal-Goldman}, {Meadows}, \&
  {Deshpande}}]{kopparapu2013}
{Kopparapu}, R.~K., {Ramirez}, R., {Kasting}, J.~F., {et~al.} 2013, \apj, 765,
  131

\bibitem[{{Leinert} {et~al.}(1998){Leinert}, {Bowyer}, {Haikala}, {Hanner},
  {Hauser}, {Levasseur-Regourd}, {Mann}, {Mattila}, {Reach}, {Schlosser},
  {Staude}, {Toller}, {Weiland}, {Weinberg}, \& {Witt}}]{leinert1998}
{Leinert}, C., {Bowyer}, S., {Haikala}, L.~K., {et~al.} 1998, \aaps, 127, 1

\bibitem[{{Lyon}(2015)}]{lyon2015}
{Lyon}, R. 2015

\bibitem[{{Mennesson} {et~al.}(2014){Mennesson}, {Millan-Gabet}, {Serabyn},
  {Colavita}, {Absil}, {Bryden}, {Wyatt}, {Danchi}, {Defr{\`e}re}, {Dor{\'e}},
  {Hinz}, {Kuchner}, {Ragland}, {Scott}, {Stapelfeldt}, {Traub}, \&
  {Woillez}}]{mennesson2014}
{Mennesson}, B., {Millan-Gabet}, R., {Serabyn}, E., {et~al.} 2014, \apj, 797,
  119

\bibitem[{{N'Diaye} {et~al.}(2015){N'Diaye}, {Pueyo}, \&
  {Soummer}}]{mamadou2015}
{N'Diaye}, M., {Pueyo}, L., \& {Soummer}, R. 2015, \apj, 799, 225

\bibitem[{{Petigura} {et~al.}(2013{\natexlab{a}}){Petigura}, {Howard}, \&
  {Marcy}}]{petigura2013_2}
{Petigura}, E.~A., {Howard}, A.~W., \& {Marcy}, G.~W. 2013{\natexlab{a}},
  Proceedings of the National Academy of Science, 110, 19273

\bibitem[{{Petigura} {et~al.}(2013{\natexlab{b}}){Petigura}, {Marcy}, \&
  {Howard}}]{petigura2013}
{Petigura}, E.~A., {Marcy}, G.~W., \& {Howard}, A.~W. 2013{\natexlab{b}}, \apj,
  770, 69

\bibitem[{{Robinson} {et~al.}(2011){Robinson}, {Meadows}, {Crisp}, {Deming},
  {A'Hearn}, {Charbonneau}, {Livengood}, {Seager}, {Barry}, {Hearty},
  {Hewagama}, {Lisse}, {McFadden}, \& {Wellnitz}}]{robinson2011}
{Robinson}, T.~D., {Meadows}, V.~S., {Crisp}, D., {et~al.} 2011, Astrobiology,
  11, 393

\bibitem[{{Savransky} {et~al.}(2010){Savransky}, {Kasdin}, \&
  {Cady}}]{savransky2010}
{Savransky}, D., {Kasdin}, N.~J., \& {Cady}, E. 2010, \pasp, 122, 401

\bibitem[{{Shaklan} {et~al.}(2013){Shaklan}, {Levine}, {Foote}, {Rodgers},
  {Underhill}, {Marchen}, \& {Klein}}]{shaklan2013}
{Shaklan}, S., {Levine}, M., {Foote}, M., {et~al.} 2013, in Society of
  Photo-Optical Instrumentation Engineers (SPIE) Conference Series, Vol. 8864,
  Society of Photo-Optical Instrumentation Engineers (SPIE) Conference Series,
  15

\bibitem[{{Silburt} {et~al.}(2014){Silburt}, {Gaidos}, \& {Wu}}]{silburt2014}
{Silburt}, A., {Gaidos}, E., \& {Wu}, Y. 2014, ArXiv e-prints

\bibitem[{{Stark} {et~al.}(2014){Stark}, {Roberge}, {Mandell}, \&
  {Robinson}}]{stark2014_2}
{Stark}, C.~C., {Roberge}, A., {Mandell}, A., \& {Robinson}, T.~D. 2014, ArXiv
  e-prints

\bibitem[{{Turnbull} {et~al.}(2012){Turnbull}, {Glassman}, {Roberge}, {Cash},
  {Noecker}, {Lo}, {Mason}, {Oakley}, \& {Bally}}]{turnbull2012}
{Turnbull}, M.~C., {Glassman}, T., {Roberge}, A., {et~al.} 2012, \pasp, 124,
  418

\bibitem[{{Weiss} \& {Marcy}(2014)}]{weiss2014}
{Weiss}, L.~M., \& {Marcy}, G.~W. 2014, \apjl, 783, L6

\bibitem[{{Wolfgang} \& {Lopez}(2014)}]{wolfgang2014}
{Wolfgang}, A., \& {Lopez}, E. 2014, ArXiv e-prints

\bibitem[{{Zahnle} \& {Catling}(2013)}]{zahnle2013}
{Zahnle}, K.~J., \& {Catling}, D.~C. 2013, in Lunar and Planetary Inst.
  Technical Report, Vol.~44, Lunar and Planetary Science Conference, 2787

\end{thebibliography}

\end{document}